\documentclass[preprints,article,accept,moreauthors,pdftex]{Definitions/mdpi}
\graphicspath{{Definitions/}}
\firstpage{1}
\pubvolume{8}
\issuenum{9}
\articlenumber{476}
\pubyear{2022}
\copyrightyear{2022}
\datereceived{3 August 2022}
\dateaccepted{7 September 2022}
\datepublished{10 September 2022}
\hreflink{https://doi.org/10.3390/\linebreak universe8090476} 

\pdfoutput=1

\Title{Astroparticle Constraints from Cosmic Reionization and Primordial Galaxy Formation}

\TitleCitation{Astroparticle Constraints from Cosmic Reionization and Primordial Galaxy Formation}

\Author{Andrea Lapi $^{1,2,3,4,}$*\orcidB{}, Tommaso Ronconi $^{1,2}$\orcidA{}, Lumen Boco $^{1,2,3}$, Francesco Shankar $^{5}$, Nicoletta Krachmalnicoff $^{1,2,3}$, Carlo Baccigalupi $^{1,2,3}$ and Luigi Danese $^{1,2}$}

\AuthorCitation{Lapi, A.; Ronconi, T.; Boco, L.; Shankar, F.; Krachmalnicoff, N.; Baccigalupi, C.; Danese, L.}

\address{$^{1}$ \quad Scuola Internazionale Superiore Studi Avanzati (SISSA), Physics Area, Via Bonomea 265, 34136 Trieste, Italy; tronconi@sissa.it (T.R.); lboco@sissa.it (L.B.);  nkrach@sissa.it (N.K.); bacci@sissa.it (C.B.); danese@sissa.it (L.D.)\\
$^{2}$ \quad IFPU---Institute for Fundamental Physics of the Universe, Via Beirut 2, 34014 Trieste, Italy\\
$^{3}$ \quad INFN-Sezione di Trieste, Via Valerio 2, 34127 Trieste,  Italy\\
$^{4}$ \quad IRA-INAF, Via Gobetti 101, 40129 Bologna, Italy\\
$^{5}$ \quad Department of Physics and Astronomy, University of Southampton, Highfield SO17 1BJ, UK; f.shankar@soton.ac.uk}

\corres{Correspondence: lapi@sissa.it}

\abstract{We derived astroparticle constraints in different dark matter scenarios that are alternatives to cold dark matter (CDM): thermal relic warm dark matter, WDM; fuzzy dark matter, $\psi$DM; self-interacting dark matter, SIDM; sterile neutrino dark matter, $\nu$DM. Our framework is based on updated determinations of the high-redshift UV luminosity functions for primordial galaxies to redshift $z\sim 10$, on redshift-dependent halo mass functions in the above DM scenarios from numerical simulations, and on robust constraints on the reionization history of the Universe from recent astrophysical and cosmological datasets. First, we built an empirical model of cosmic reionization characterized by two parameters, namely the escape fraction $f_{\rm esc}$ of ionizing photons from primordial galaxies, and the limiting UV magnitude $M_{\rm UV}^{\rm lim}$ down to which the extrapolated UV luminosity functions steeply increased. Second, we performed standard abundance matching of the UV luminosity function and the halo mass function, obtaining a relationship between UV luminosity and the halo mass, whose shape depends on an astroparticle quantity $X$ specific to each DM scenario (e.g., WDM particle mass); we exploited such a relationship to introduce (in the analysis) a constraint from primordial galaxy formation, in terms of the threshold halo mass above which primordial galaxies can efficiently form stars. Third, we performed Bayesian inference on the three parameters $f_{\rm esc}$, $M_{\rm UV}^{\rm lim}$, and $X$ via a standard MCMC technique, and compared the outcomes of different DM scenarios on the reionization history. We also investigated the robustness of our findings against educated variations of still uncertain astrophysical quantities. Finally, we highlight the relevance of our astroparticle estimates in predicting the behavior of the high-redshift UV luminosity function at faint, yet unexplored magnitudes, which may be tested with the advent of the James Webb Space Telescope.}


\begin{document}

\section{Introduction}\label{sec|intro}

Many astrophysical probes and cosmological experiments have firmly established that most of the matter content of the Universe is dark, i.e., is constituted by particles suffering very weak or negligible interactions with baryons apart from long-range gravitational forces. However, so far, such dark matter (DM) particles have escaped firm detection, both in colliders \cite{Mitsou13,Kahlhoefer17,Argyropoulo21} and from direct \cite{Aprile18,Bernabei20} or indirect \cite{Fermi15,Ackermann17,Albert17,Zornoza21} searches in the sky.

The standard paradigm envisages DM to be constituted by weakly interacting particles (such as supersymmetric neutralinos or gravitinos) with masses in the GeV range \cite{Bertone18}. Such a form of DM is said to be cold, meaning that particles are non-relativistic at the epoch of decoupling and feature negligible free-streaming\endnote{Free-streaming is the process through which small-scale perturbations can be erased if particles with residual thermal velocities diffuse out of them before collapse.}. As a consequence, bound cold DM (CDM) structures, called halos, grow sequentially in time and hierarchically in mass by stochastically merging together \cite{Frenk12,Lapi20}.

On large, cosmological scales, such a picture is remarkably consistent with the data, and most noticeably with microwave background detection experiments \cite{Planck20}. However, on galactic and subgalactic scales, the CDM hypothesis has been challenged by various issues, including: the flat shape of the inner density profiles in DM-dominated dwarfs with respect to the steep behavior measured in the halos of $N$-body simulations \cite{Navarro97,deBlok08}; the discrepancy between the number and dynamical properties of observed Milky Way satellites with respect to those of subhalos in gravity-only simulations \cite{Boylan12,Bullock17}; the emergence of tight relationships between properties of the dark and luminous components in disc-dominated galaxies, such as the universal core surface density or the radial acceleration relation \cite{Gentile09,McGaugh16}, which may be indicative of a new dark sector and/or of non-gravitational coupling between DM particles and baryons. One possible explanation for the above effects invokes physical processes that can cause violent fluctuations in the inner gravitational potential and/or the  transfer of energy and angular momentum from the baryons to DM, such as dynamical friction \cite{ElZant01,Tonini06} or feedback effects from stars and active galactic nuclei~\cite{Pontzen14,Peirani17,Freundlich20}.

An alternative, perhaps more fascinating solution is to abandon the CDM hypothesis and look at nonstandard particle candidates \cite{Bertone04,Feng10,Salucci21}. A few examples often considered in the literature and relevant for the present paper include: thermal warm dark matter (WDM) relics with masses $\sim$ a few keVs \cite{Bode01,Viel13,Lovell14}; fuzzy or particle-wave dark matter ($\psi$DM), i.e., Bose–Einstein condensates of ultralight axions with masses $\gtrsim 10^{-22}$ eV \cite{Hu00,Hui17}; self-interacting dark matter (SIDM) mediated by a massive dark photon decaying to a light--dark fermion \cite{Vogelsberger16,Tulin18,Huo18}; non-thermally produced sterile neutrinos dark matter ($\nu$DM) with the keV-scale mass, and given lepton asymmetry \cite{Seljak06,Kusenko09,Adhikari17}. As a consequence of free-streaming, quantum pressure effects, and/or dark-sector interaction, all these scenarios produce a matter power spectrum suppressed on small scales, fewer (sub)structures, and flatter inner density profiles within halos relative to CDM \cite{Schneider13,Vogelsberger16,Dayal15,Schive16,Huo18,Menci18,Lovell20,Romanello21,Kulkarni22}.
Indirect astrophysical constraints on the properties of such nonstandard DM scenarios, and especially of thermal WDM relics, have been obtained by investigating the Lyman-$\alpha$ forest \cite{Viel13,Irsic17wdm,Irsic17fdm}, high-redshift galaxy counts \cite{Pacucci13,Menci16,Shirasaki21,Sabti22}, $\gamma$-ray bursts \cite{deSouza12,Lapi17}, cosmic reionization \cite{Barkana01,Lapi15,Dayal17,Carucci19}, integrated 21cm data \cite{Carucci15,Boyarsky19,Chatterjee19,Rudakovskyi20}, $\gamma$-ray emission \cite{Bringmann17,Grand22}, fossil records of the Local Group \cite{Weisz14,Weisz17}, and Milky Way satellite galaxies \cite{Kennedy14,Nadler21,Newton21}.

The present paper will mainly focus on the constraints to DM that can be derived from cosmic reionization. This is the process by which the intergalactic medium has transitioned again to an ionized state (it was already fully ionized before the epoch of recombination, when the Universe was younger than 380,000 years) due to the radiation from the first astrophysical sources, such as primordial galaxies and AGNs. Reionization constitutes a natural bridge between galaxy formation and the underlying cosmological model; in a nutshell, the basic argument runs as follows. The history of cosmic reionization, as reconstructed from cosmological and astrophysical observations, can be exploited to gauge the level of the ionizing background from primordial galaxies, and in turn (although with some assumptions to be discussed next) to probe their number densities. Such galaxies are faint and tend to live within small halos so that their numbers can inform us about the shape of the halo mass distribution and of the power spectrum at the low mass end, which is sensitive to the microscopic properties of the DM particles.

More specifically, in the present paper, we aimed to obtain revisited and additional astroparticle constraints for the aforementioned DM scenarios (WDM, $\psi$DM, SIDM, and $\nu$DM) by combining different ingredients: (i) updated measurements of the high-redshift UV luminosity functions for primordial galaxies out to redshift $z\sim 10$; (ii) precision determination of the redshift-dependent halo mass functions in different DM scenarios from numerical simulations; (iii) recent constraints on the reionization history of the Universe from astrophysical and cosmological probes. Such ingredients were exploited to build an empirical model of cosmic reionization to be compared with the data. The model  depended on three basic parameters: the escape fraction of ionizing photons from primordial galaxies, the limiting UV magnitude down to which the UV luminosity function steeply increased, and an astroparticle property specific to the DM scenario (e.g., WDM particle mass). We then performed Bayesian inference on these three parameters via a standard MCMC technique, and at the same time, we investigated how the astroparticle constraints were degenerated with (and robust against) variations in crucial (but still uncertain) astrophysical quantities. The structure of the paper is as follows: in Section \ref{sec|methods} we describe our methods and analysis; in Section \ref{sec|results} we present and discuss our results; in Section \ref{sec|summary}, we summarize our findings and future perspectives. Throughout the work, we adopted the standard, flat cosmology \cite{Planck20} with rounded parameter values: matter density $\Omega_M \approx 0.31$, baryon density $\Omega_b \approx 0.05$, Hubble constant $H_0 = 100\, \rm{h}$ km s$^{-1}$ Mpc$^{-1}$ with $h\approx 0.68$. A Chabrier \cite{Chabrier03} initial mass function was assumed.

\section{Methods and Analysis}\label{sec|methods}

In this section, we present our empirical model of reionization, derive a galaxy formation constraint from the abundance matching of the luminosity functions with the halo mass functions in a specific DM scenario, and finally describe our estimation procedure based on a Bayesian MCMC technique.

\subsection{An Empirical Model of Reionization}\label{sec|reion}

\begin{figure}[H]

\includegraphics[width=9.5cm]{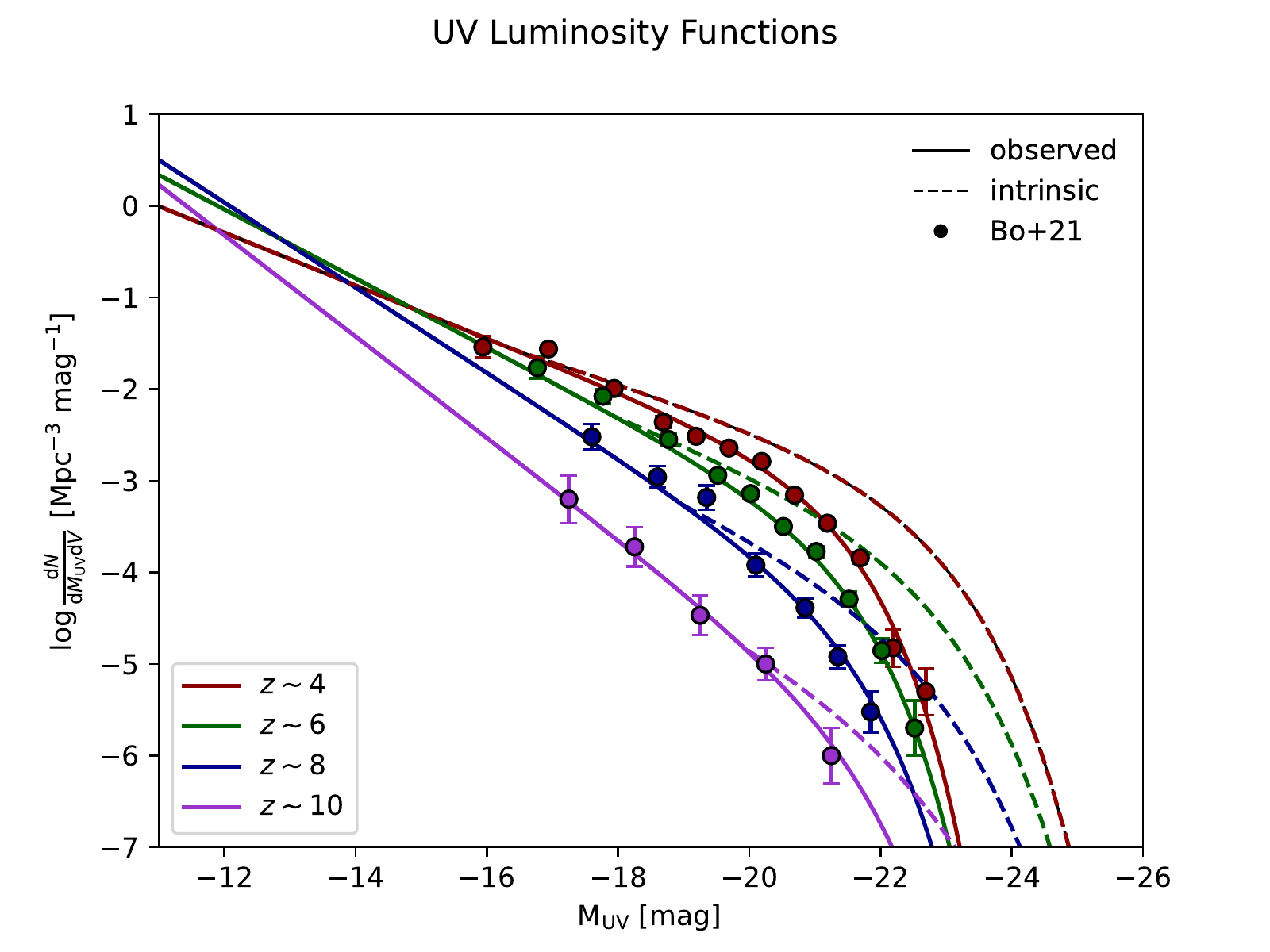}
\caption{{The} UV luminosity functions at redshifts $z\sim 4$ (red), $6$ (green), $8$ (blue), and $10$ (purple). Data points (circles) and fits are from \cite{Bouwens21,Oesch18}. Solid lines illustrate the observed luminosity functions, while dashed lines illustrate the intrinsic ones, after correction for dust extinction via the UV continuum slope according to the procedure by \cite{Bouwens14}.}\label{fig|UVLF}
\end{figure}

To build a simple empirical model of reionization, we start from the recent determination of the UV luminosity functions by \cite{Bouwens21,Oesch18} out to redshift $z\sim 10$. Specifically, in Figure~\ref{fig|UVLF} we illustrate the binned luminosity functions (filled circles) at $\approx 1600$ {\AA} in the relevant redshift range $z\sim 4$--10 (color-coded), together with the corresponding continuous Schechter function rendition (solid lines) in the form
${\rm d}N/{\rm d}M_{\rm UV}\,{\rm d}V \propto 10^{-0.4\,(M_{\rm UV}-M_{\rm UV}^\star)\,(\alpha+1)}\times \exp[-10^{-0.4\,(M_{\rm UV}-M_{\rm
UV}^\star)}]$. The luminosity functions were well determined down to a UV magnitude $M_{\rm UV}\approx -17$, with the faint end progressively steepening from a slope $\alpha\approx -1.7$ at $z\approx 4$ to $\alpha\approx -2.4$ at $z\approx 10$, and a characteristic magnitude $M_{\rm UV}^\star\approx -21$ is almost independent of redshift for $z\gtrsim 4$. Note that the UV magnitude can be related to the monochromatic UV luminosity at $1600$ {\AA} by the relation $\log L_{\rm UV}$ \mbox{[erg s$^{-1}$ Hz$^{-1}$]} $\approx -0.4\, (M_{\rm UV}-51.6)$.

In Figure~\ref{fig|UVLF}, we also report the intrinsic luminosity functions after correction for dust extinction (dashed lines), which have been computed exploiting the relation between extinction, the slope of the UV spectrum, and observed UV magnitude by \cite{Meurer99,Bouwens14}; we caveat the reader that such a dust correction can be considered well-established and robust only for UV magnitude $M_{\rm UV}\gtrsim -22$ and $z\lesssim 8$. The figure shows that the effects of dust extinction on the galaxy statistics are negligible in the present context since cosmic reionization is majorly contributed by faint galaxies with $M_{\rm UV}\gtrsim -17$, where the intrinsic and observed luminosity functions are practically indistinguishable.
The intrinsic UV luminosity is routinely linked to the physical star formation rate (SFR) of galaxies by the relation $L_{\rm UV} = \kappa_{\rm UV}~\times~$SFR, with the quantity $\kappa_{\rm UV}$ depending somewhat on the IMF, on galactic age and on chemical composition (see \cite{Kennicutt12,Madau14,Cai14,Robertson15,Finkelstein19}); we adopt as a reference value $\kappa_{\rm UV}\approx 1.5\times 10^{28}$ erg s$^{-1}$ Hz$^{-1}$ $\rm{M}_{\odot}^{-1}$ year, apt for a Chabrier IMF, age $\gtrsim 10^8$ years, and appreciably sub-solar metallicity. Then the relation $\log$ SFR [M$_\odot$ year$^{-1}$] $\approx -0.4\,(M_{\rm UV}+18.5)$~holds.

From the intrinsic UV luminosity functions, we compute the cosmic SFR density as
\begin{equation}
\rho_{\rm SFR}(z)=\int^{M_{\rm UV}^{\rm lim}}{\rm d}M_{\rm UV}\, \frac{{\rm d}N}{{\rm d}M_{\rm UV}\,{\rm d}V}\, {\rm SFR}\; ,
\end{equation}
where $M_{\rm UV}^{\rm lim}$ represents a limiting magnitude down to which the luminosity function is steeply increasing; the rationale is that at magnitudes fainter than such a threshold, the luminosity function bends downwards because the galaxy formation process becomes inefficient and/or because the power spectrum is cut-off due to the microscopic nature of DM \cite{Bose18,Romanello21,Munoz22}.~The quantity $M_{\rm UV}^{\rm lim}$ is uncertain since the observations of the UV luminosity function are limited to $M_{\rm UV}\approx -17$; thus, it will be treated as a free parameter in our Bayesian analysis discussed in Section \ref{sec|Bayes}.

Then we compute the cosmic ionization photon rate as
\begin{equation}\label{eq|Ndotion}
\dot N_{\rm ion}\approx f_{\rm esc}\, k_{\rm ion} \, \rho_{\rm SFR} + \dot N_{\rm ion}^{\rm AGN} \;;
\end{equation}
here $k_{\rm ion}\approx 4\times 10^{53}$ is the number of ionizing photons s$^{-1}$ M$_\odot^{-1}$ year appropriate for a Chabrier IMF, $f_{\rm esc}$ is the average escape fraction of ionizing photons from primordial galaxies \cite{Mao07,Cai14,Robertson15,Rutkowski16}, and $\dot N_{\rm ion}^{\rm AGN}$ is the contribution to the ionization rate from active galactic nuclei (AGNs).

The escape fraction from primordial galaxies is still a very uncertain quantity, with estimates ranging from a few percentage points to a few tens of percentage points \cite{Paardekooper15,Vanzella18,Alavi20,Smith20,Izotov21,Pahl21,Atek22,Naidu22}. We will keep $f_{\rm esc}$ as a free parameter in our Bayesian analysis of Section \ref{sec|Bayes}, to highlight the impact of such an astrophysical uncertainty on the astroparticle constraints. However, we will also explore the implication of adopting a redshift-dependent escape fraction $f_{\rm esc}(z)$ increasing from small values $5\%$ in the local Universe to around $\approx$20\% at a high redshift, as suggested by cosmological radiative transfer simulations of the UV background \cite{Puchwein19}.

The quantity $k_{\rm ion}$ entering in Equation (\ref{eq|Ndotion}) is also somewhat uncertain because it depends on the adopted IMF, metallicity, and other stellar population properties; however, its values have been shown not to vary wildly \cite{Mao07,Cai14,Bouwens14,Lapi15,Finkelstein19}. Perhaps the only exception is when a hypothetically strongly top-heavy IMF is assumed, since in that case the number of ionizing photons is considerably enhanced; however, note that such an IMF would imply a correspondingly stronger metal and dust enrichment of the interstellar medium in primordial galaxies already at $z\gtrsim 8$, which is not expected for these faint UV sources and in turn, would dramatically reduce their ionization efficiency. For the sake of simplicity, hereafter we will assume the aforementioned definite value of $k_{\rm ion}$, but notice that any constraint derived on $f_{\rm esc}$ should actually be referred to as the combined quantity $f_{\rm esc}\times (k_{\rm ion}/4\times 10^{53}\; \mathrm{s}^{-1}\, \mathrm{M}_\odot^{-1}\, \mathrm{year})$.

As for the AGN contribution $\dot N_{\rm ion}^{\rm AGN}$ appearing in Equation (\ref{eq|Ndotion}), we adopt the redshift-dependent parameterization by \cite{Shen20}
\begin{equation}\label{eq|Ndotion_AGN}
\dot N_{\rm ion}^{\rm AGN}\approx 1.1\times 10^{50}\, f_{\rm esc}^{\rm AGN} \frac{(1+z)^{5.865}\, e^{0.731\,z}}{15.6+e^{3.055\,z}}\; ,
\end{equation}
in units of ionizing photons s$^{-1}$ Mpc$^{-3}$, which is based on the latest determination of the bolometric AGN luminosity functions. These imply a rapid decline in the number density of bright quasars and a relative paucity of faint AGNs for $z\gtrsim 4$, though the latter point is still somewhat debated \cite{Shankar07,Giallongo15,Ricci2017,Giallongo19,Kulkarni19,Ananna20,Shen20}. As a consequence, the AGN contribution to the overall $\dot N_{\rm ion}$ for $z\gtrsim 5$ is minor with respect to primordial galaxies, even when escape fractions around $f_{\rm esc}^{\rm AGN}\sim 100\%$ are adopted \cite{Grazian18,Romano19}.

We then exploit $\dot N_{\rm ion}(z)$ from Equation (\ref{eq|Ndotion}) to compute the hydrogen ionizing fraction $Q_{\rm HII}$ from the evolution equation
\begin{equation}\label{eq|QHII}
\dot Q_{\rm HII} = \frac{\dot N_{\rm ion}}{\bar n_{\rm H}}-\frac{Q_{\rm HII}}{
t_{\rm rec}}\; ,
\end{equation}
that takes into account the competition between ionization and recombination
processes~\cite{Madau99,Loeb01,Mao07,Cai14}. In the above, $\bar n_{\rm
H}\approx 2\times 10^{-7}\, (\Omega_b h^2/0.022)$ cm$^{-3}$ is the mean
co-moving hydrogen number density, and $t_{\rm rec}\approx 3.2$ Gyr $[(1+z)/7]^{-3}\, C_{\rm HII}^{-1}$ is the recombination timescale for the case B coefficient and an IGM temperature of $2\times 10^4$ K. The quantity $C_{\rm HII}$ (appearing in the recombination time) is the clumping factor of the ionized hydrogen, for which we adopt the redshift-dependent parameterization $C_{\rm HII} \approx {\rm min}[1+43\,z^{-1.71},20]$ \mbox{by \cite{Pawlik09,Haardt12}.}

Finally, we compute the electron scattering optical depth out to redshift $z$ from
\begin{equation}\label{eq|tau_es}
\tau_{\rm es}(<z) = c\, \sigma_{\rm T}\,\bar n_{\rm H}\int^z{\rm d}z'\,f_e\,Q_{\rm HII}(z')
(1+z')^2 \, H^{-1}(z')\; ,
\end{equation}
where $H(z)=H_0\,[\Omega_M\,(1+z)^3+1-\Omega_M]^{1/2}$ is the Hubble
parameter, $c$ is the speed of light, $\sigma_{\rm T}$ the Thomson cross
section, and $f_e\approx 1+\eta\, Y/4\,X$ is the number of free-electron;
we adopt primordial abundances $Y\approx 0.2454$ and $X\approx 1-Y$, and complete double helium ionization at $z\sim 4$ so that $\eta\approx 2$ for $z\lesssim 4$ and $\eta\approx 1$ for $z\gtrsim 4$.

\subsection{A Constraint from Primordial Galaxy Formation}\label{sec|abma}

As mentioned in Section \ref{sec|intro}, we consider different DM scenarios alternative to CDM: thermal warm dark matter (WDM) relics; fuzzy dark matter ($\psi$DM); self-interacting dark matter (SIDM); sterile neutrinos dark matter ($\nu$DM). In all these scenarios, the number of small-mass halos is reduced relative to CDM; this is best specified in terms of the halo mass function, namely the number density of halos per co-moving volume and halo mass $M_{\rm H}$ bins, which can be conveniently written in terms of the CDM one as
\begin{equation}\label{eq|HMF}
\frac{{\rm d}N_X}{{\rm d}M_{\rm H}\, {\rm d}V}= \frac{{\rm d}N_{\rm CDM}}{{\rm d} M_{\rm H}\, {\rm d}V}\, \left[1+\left(\alpha\, \frac{M_{\rm H}^{\rm cut}}{M_{\rm H}}\right)^\beta\right]^{-\gamma}\;,
\end{equation}
where $\alpha$, $\beta$ and $\gamma$ are shape parameters, and $M_{\rm H}^{\rm cut}$ is a cutoff halo mass. We compute the CDM halo mass function by exploiting the Python \texttt{COLOSSUS} package \cite{Diemer18} and the fitting formula by \cite{Tinker08} for virial masses. The parameters $(\alpha,\beta,\gamma)$ in Equation (\ref{eq|HMF}) are instead derived from fits to the outcomes of numerical simulations in the considered DM scenarios; the related values of the parameters, and the literature works from which these are taken, are reported in Table \ref{tab|HMFpar}. We stress that for deriving robust constraints on different DM scenarios based on the halo mass function it is extremely important to rely on the results from detailed simulations (as done here), and not on semi-analytic derivations based on the excursion set formalism, whose outcomes on the shape of the mass function for masses $M_{\rm H}\lesssim M_{\rm H}^{\rm cut}$ are rather sensitive to several assumptions (e.g., the  filter function used in deriving the mass variance from the power spectrum, the mass-dependence in the collapse barrier, etc.; e.g., \cite{Schneider13,Lapi15}).

\begin{table}[H]
\caption{Parameters describing the ratio of the halo mass function for different DM scenarios relative to the standard CDM in terms of the expression $[1+(\alpha\, M_{\rm H}^{\rm cut}/M_{\rm H})^\beta]^{-\gamma}$, where $M_{\rm H}$ is the halo mass and $M_{\rm H}^{\rm cut}$ is a characteristic cutoff scale, see Section \ref{sec|abma} for details. The values of the parameters $(\alpha,\beta,\gamma)$, extracted from fits to the outcomes of numerical simulations in the considered DM scenarios, are taken from the literature studies referenced in the last column.}\label{tab|HMFpar}
\newcolumntype{C}{>{\centering\arraybackslash}X}
\begin{tabularx}{\textwidth}{CCCCC}
\toprule
\textbf{Scenario} & \boldmath{$\alpha$} &\boldmath{ $\beta$} & \boldmath{$\gamma$ }& \textbf{Ref.}\\
\midrule
WDM & $1.0$ &  $1.0$ & $1.16$ & \cite{Schneider12}\\
$\psi$DM & $1.0$ &  $1.1$ & $2.2$ & \cite{Schive16}\\
SIDM & $1.0$ &  $1.0$ & $1.34$ & \cite{Huo18}\\
$\nu$DM & $2.3$ &  $0.8$ & $1$ & \cite{Lovell20}\\
\bottomrule
\end{tabularx}

\end{table}

As to the cutoff mass $M_{\rm H}^{\rm cut}$, in WDM it is determined by free-streaming effects \cite{Schneider12} and reads $M_{\rm H}^{\rm cut} \approx 1.9\times 10^{10}$ M$_\odot\, (m_X/{\rm keV})^{-3.33}$ in terms of the particle mass $m_X$. The quantity $M_{\rm H}^{\rm cut}$ for WDM is also referred to as the half-mode mass, representing the mass where the amplitude of the WDM transfer function, i.e., the square root of the ratio between the WDM and the CDM power spectra, is reduced to 50\%.~Note that this is substantially larger (a factor of a few $10^3$) than the free streaming mass, i.e., the mass related to the typical length-scale for diffusion of WDM particles out of primordial perturbations. In $\psi$DM, $M_{\rm H}^{\rm cut} \approx 1.6\times 10^{10}$ M$_\odot$ $\, (m_X/10^{-22}\, {\rm eV})^{-1.33}$ is related to the coherent behavior of the condensate \cite{Schive16} for axions with mass $m_X$. In the SIDM scenario, $M_{\rm H}^{\rm cut}\approx 7\times 10^7$ M$_\odot\, (T_X/{\rm keV})^{-3}$ can be linked to the visible sector temperature $T_X$ when kinetic decoupling of the DM particles takes place \cite{Huo18}. In the $\nu$DM scenario, $M_{\rm H}^{\rm cut}$ depends not only on the particle mass $m_X$ but also on the lepton asymmetry $L_X$ with which sterile neutrinos are generated out of thermal equilibrium in the early Universe \cite{Lovell20}. In this work, we actually set the sterile neutrino mass to $m_X\approx 7$ keV since such a particle may constitute an interesting candidate to explain the $3.55$ keV line detected in stacked X-rays observations of galaxy clusters \cite{Bulbul14}. The corresponding values of the cutoff mass as a function of $L_X$ are non-monotonic, starting from $M_{\rm H}^{\rm cut}\approx 2.4\times 10^9$ M$_\odot$ for $L_X\approx 1$, then decreasing to a minimum $\approx 1.3\times 10^8$ M$_\odot$ for $L_X\approx 8$ and then increasing again to $\approx 9.2\times 10^8$ M$_\odot$ for $L_X\approx 11$ up to $3.1\times 10^9$ M$_\odot$ for the maximal $L_X\approx 120$.

\begin{figure}[H]

\includegraphics[width=12cm]{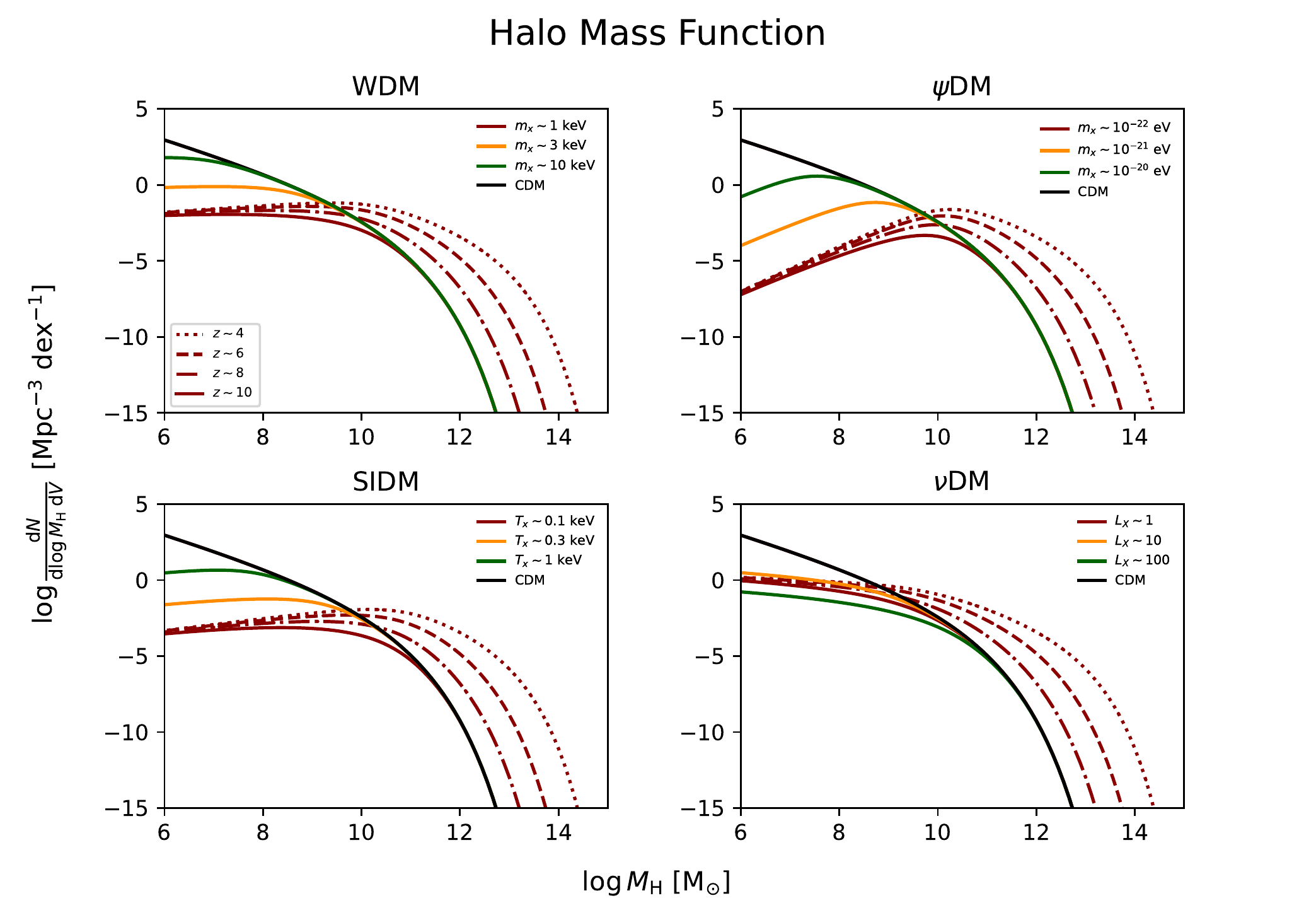}
\caption{The halo mass function in different DM scenarios.
Different line styles illustrate the evolution with redshift (only plotted for one value of the astroparticle property), as reported in the legend. Different colors illustrate the change in the mass function at $z\sim 10$ as a function of the astroparticle property, as detailed in the legend; for reference, the black line refers to the \mbox{standard CDM.}}\label{fig|HMF}
\end{figure}

In Figure~\ref{fig|HMF}, we illustrate the halo mass functions in the different DM scenarios, to highlight the dependence on redshift and the particle property. For example, focusing on WDM, it is seen that at a given redshift $z\sim 10$ (solid lines with different colors) the halo mass function progressively flattens with respect to that in standard CDM (black line); the deviation occurs at smaller halo masses for higher WDM particle masses $m_X$, so that the CDM behavior is recovered for $m_X\rightarrow \infty$. At a given particle mass, $m_X\sim 1$ keV (red lines with different line styles), the exponential cutoff of the mass function shifts to larger halo masses for decreasing redshift, reflecting the hierarchical clustering of halos. In the other DM scenarios, the behavior is similar, but the shape of the mass function past the low-mass end flattening can be appreciably different; e.g., in the $\psi$DM scenario the mass function is strongly suppressed for small masses and actually bends downward rather than flattening, implying a strong reduction or even an absence of low mass halos.

\begin{figure}[H]

\includegraphics[width=12cm]{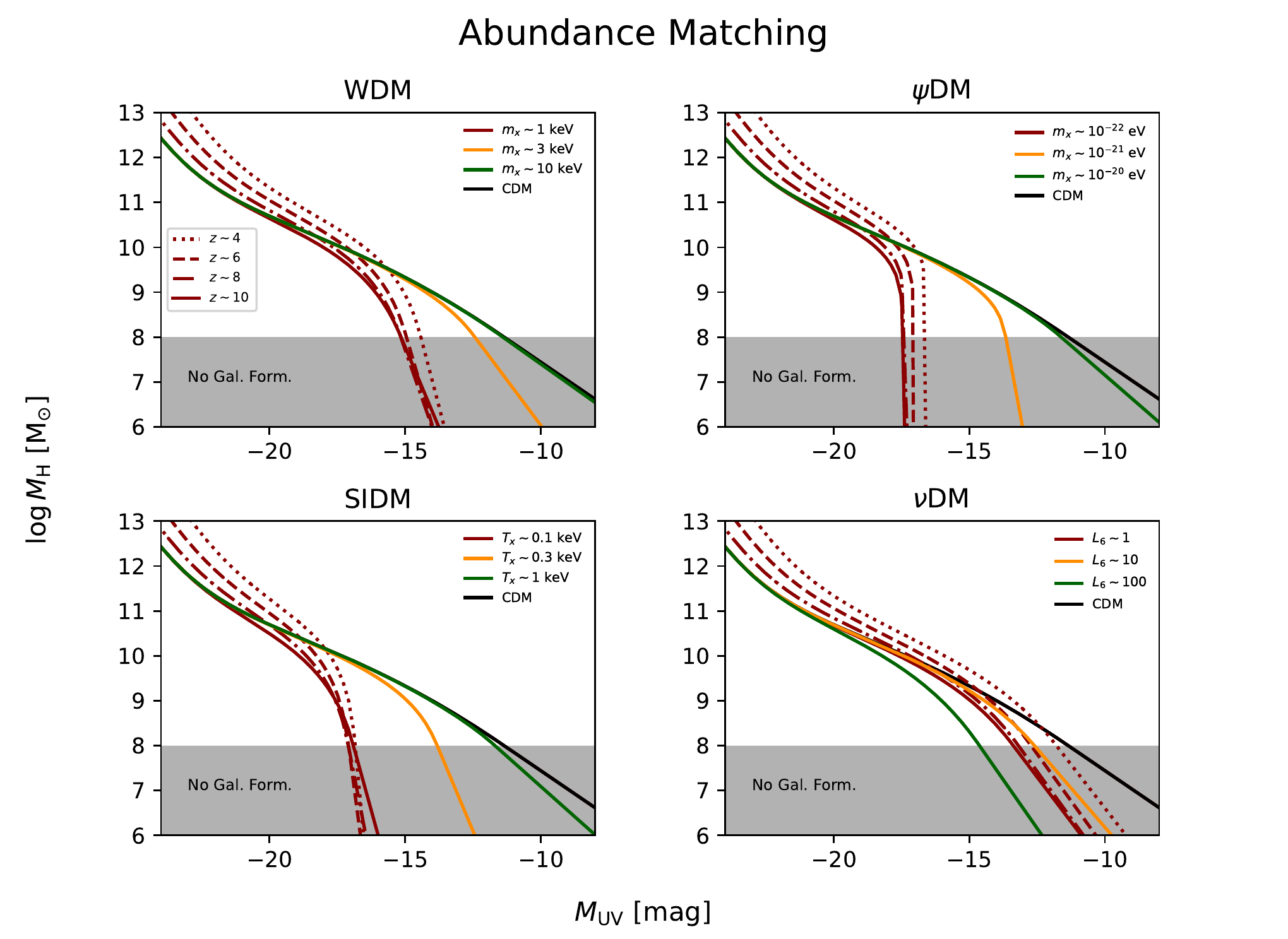}
\caption{Relationship between the halo mass $M_{\rm H}$ and the UV magnitude $M_{\rm UV}$, derived from the abundance matching of the observed UV luminosity function and the halo mass function (see text for details) in different DM scenarios. Different line styles illustrate the evolution with redshift (only plotted for one value of the astroparticle property), as reported in the legend. Different colors illustrate the change in the mass function at $z\sim 10$  when varying the astroparticle property, as detailed in the legend; for reference, the black line refers to the standard CDM. In all panels, the grey shaded area marks the region below the threshold halo mass $M_{\rm H}^{\rm GF}$ where primordial galaxy formation becomes inefficient (see Section \ref{sec|abma}).}\label{fig|abma}
\end{figure}

We are now left with an observed UV luminosity function that is required to be steep down to $M_{\rm UV}^{\rm lim}$ for reproducing the reionization history, and at the same time a halo mass function that progressively flattens or even bends down for halo masses smaller than $M_{\rm H}^{\rm cut}$. This necessarily implies that the relationship between UV magnitude and halo masses must differ from the CDM case and depend on the DM scenario. Such a relationship may be derived via the standard abundance matching technique \cite{Aversa15,Moster18,Cristofari19,Behroozi20}, i.e., matching the cumulative number densities in galaxies and halos according to the expression
\begin{equation}\label{eq|abma}
\int_{M_{\rm H}}^{+\infty}{\rm d}M_{\rm H}'\;\frac{{\rm d}N_X}{{\rm d}M_{\rm H}'\, {\rm d}V}(M_{\rm H}',z|X) = \int_{-\infty}^{M_{\rm UV}}{\rm d}M_{\rm UV}'\;\frac{{\rm d}N}{{\rm d}M_{\rm UV}'\, {\rm d}V}(M_{\rm UV}',z)
\end{equation}
which implicitly defines a one-to-one monotonic relationship $M_{\rm H}(M_{\rm UV},z|X)$; here the quantity $X$ stands for the specific property of the DM scenario that determines the behavior of the mass function for $M_{\rm H}\lesssim M_{\rm H}^{\rm cut}$: particle mass $m_X$ for WDM and $\psi$DM, kinetic temperature $T_X$ for SIDM, and lepton asymmetry $L_X$ for $\nu$DM.  In Figure~\ref{fig|abma}, we show the outcome of this procedure in the different DM scenarios, highlighting its dependence on redshift and the particle property. Focusing on WDM as a representative case, it is seen that at a given redshift $z\sim 10$ (solid lines with different colors) the $M_{\rm H}(M_{\rm UV}|X=m_X)$ relation progressively steepens with respect to the standard CDM case (black line), and more for smaller $m_X$; at the other end, the relation becomes indistinguishable from that in CDM for particle masses $m_X\gtrsim$ some keVs. At a given particle mass $m_X\sim 1$ keV (red lines with different line styles), the relation $M_{\rm H}(M_{\rm UV},z|m_X)$ barely depends on redshift, at least in the range $z\sim 4$--10 relevant to this work, because the cosmic evolution of the UV luminosity function and the halo mass function mirror each other (see discussion by \cite{Bouwens21}). In the other DM scenarios, the behavior of the $M_{\rm H}(M_{\rm UV},z|X)$ relation is similar but its shape for small halo masses is appreciably different; e.g., in the $\psi$DM scenario, the relation is substantially steeper, reflecting the paucity of small halos in the mass function (see above)

In all panels of Figure~\ref{fig|abma}, the grey shaded area illustrates the region where the galaxy formation is thought to become inefficient because of various processes \cite{Efstathiou92,Sobacchi13,Cai14,Finkelstein19}: molecular cooling may be hindered and atomic cooling may be limited given the low metallicities expected at high redshift; SN feedback can easily quench star formation in low-mass halos; star formation may be photo-suppressed by the intense diffuse UV background; the formation of massive stars at low metallicities may originate additional radiative feedback processes; etc. This inefficiency in galaxy formation is thought to occur for halo masses smaller than a critical value $M_{\rm H}^{\rm GF}\lesssim$ a few $10^8$ M$_\odot$ possibly dependent on redshift, albeit with some uncertainties due to detailed modeling of the above processes. Remarkably, it has also been pointed out that such a threshold can alleviate the missing satellite problem, because the number density of small mass halos where galaxy formation can take place becomes closer to the number of visible satellites in the Milky Way \cite{Bullock17,Lapi17}.

\textls[-15]{We conservatively adopt a threshold $M_{\rm H}^{\rm GF}\approx 10^8$ M$_\odot$ that is typically associated to photo-suppression of galaxy formation due to the UV background \cite{Finkelstein19}; other choices will be explored in Section \ref{sec|results}.~We can now introduce the self-consistency galaxy formation constraint}
\begin{equation}\label{eq|MnoGF}
M_{\rm H}(M_{\rm UV}^{\rm lim},z|X) \approx M_{\rm H}^{\rm GF}
\end{equation}
i.e., the limiting UV magnitude down to which the UV luminosity function is steeply increasing must correspond, in the given DM scenario, to the halo mass $M_{\rm H}^{\rm GF}$ (see also Section \ref{sec|reion}); in other words, for halo masses $M_{\rm H}\lesssim M_{\rm H}^{\rm GF}$, galaxy formation is hindered, and this will imply that at magnitudes fainter than $M_{\rm UV}^{\rm lim}$ the UV luminosity function will no rise any longer. We allow for a $0.25$ dex dispersion around Equation (\ref{eq|MnoGF}); this quantitatively includes both the scatter in the abundance matching relation $M_{\rm H}(M_{\rm UV}^{\rm lim},z|X)$ associated to the uncertainty in the UV luminosity functions determination (see also \cite{Aversa15}), \textls[-20]{and the theoretical uncertainty in the threshold halo mass $M_{\rm H}^{\rm GF}$ for galaxy formation (see~\cite{Cai14,Finkelstein19}).}

Note that the abundance matching procedure in Equation (\ref{eq|abma}) automatically guarantees to have, at any given redshift, the same number of halos hosting galaxies and of galaxies producing ionizing photons; in other words, the cumulative number of halos obtained integrating the halo mass function down to $M_{\rm H}^{\rm GF}$ is approximately equal to the cumulative number of galaxies obtained by integrating the UV luminosity function down to $M_{\rm UV}^{\rm lim}$. Other investigations in the literature (e.g., \cite{Carucci15,Lapi15,Menci16,Carucci19}) have adopted conditions less restricting than Equation (\ref{eq|abma}),
for example by requiring just to have a larger number of halos hosting galaxies than of galaxies contributing to cosmic reionization.

\subsection{Bayesian Analysis}\label{sec|Bayes}

The descriptions provided in the previous sections highlight that three basic parameters enter our framework: the escape fraction $f_{\rm esc}$ of ionizing photons from primordial galaxies, the limiting UV magnitude $M_{\rm UV}^{\rm lim}$ down to which the UV luminosity function is steeply increasing, and a quantity $X$ specific to the DM scenario. To estimate such parameters, we adopted a Bayesian MCMC framework, numerically implemented via the Python package \texttt{emcee} \cite{Foreman13}. We used a standard Gaussian likelihood $\mathcal{L}(\theta)\equiv -\sum_i \chi_i^2(\theta)/2$
where $\theta=\{f_{\rm esc},M_{\rm UV}^{\rm lim},m_X\}$ is the vector of parameters, and the summation is over different observables; for the latter, the corresponding $\chi_i^2= \sum_j [\mathcal{M}(z_j,\theta)-\mathcal{D}(z_j)]^2/\sigma_{\mathcal{D}}^2(z_j)$ is obtained by comparing our empirical model expectations $\mathcal{M}(z_j,\theta)$ to the data $\mathcal{D}(z_j)$ with their uncertainties $\sigma_{\mathcal{D}}^2(z_j)$, summing over the different redshifts $z_j$ of the datapoints (when~applicable).

Our overall data sample is constituted by (see summary in Table \ref{tab|data}): robust observational measurements at $z\gtrsim 4$ of the ionizing photon rate \cite{Becker13,Becker21}; constraints on the volume filling factor of ionized hydrogen provided by various astrophysical probes, including Lyman-$\alpha$ emitters and Lyman-break galaxies luminosity functions, Lyman-$\alpha$ forest dark pixels, and QSO damping wings \cite{Konno14,McGreer15,Davies18,Mason18,Konno18,Hoag19,Bolan22,Greig22}; latest constraints on the electron scattering optical depth provided by the observations of the cosmic microwave background from the \textit{Planck} \cite{Planck20} mission. We also include the galaxy formation constraint provided by a $\chi_{\rm GF}^2\sim \sum_{j}\,[M_{\rm H}(M_{\rm UV}^{\rm lim},z_j|X) - M_{\rm H}^{\rm GF}]^2/\sigma_{\rm GF}^2$, where we take $z_j=\{4,6,8,10\}$ as reference redshifts (to avoid extrapolation at redshifts where the UV luminosity functions are not well constrained) and $\sigma_{\rm GF}\approx 0.25$ dex as the typical uncertainty in the galaxy \mbox{formation~constraint.}

\begin{table}[H]\tablesize{\footnotesize}
\caption{Overview of the data considered in the Bayesian analysis of this work, referring to the ionizing photon rate $\dot N_{\rm ion}$, volume filling fraction $Q_{\rm HII}$ of ionized hydrogen, and optical depth $\tau_{\rm es}$ for electron scattering.}\label{tab|data}

\newcolumntype{C}{>{\centering\arraybackslash}X}
\begin{tabularx}{\textwidth}{p{3cm}<{\centering} Cp{2.5cm}<{\centering}CC}
\toprule
\textbf{Observable [units]} & \textbf{Redshifts} & \textbf{Values} & \textbf{Errors} & \textbf{Ref.}\\
\midrule
$\log \dot N_{\rm ion}$ $[$ s$^{-1}$ Mpc$^{-3}]$&&&&\\
& $\{4.0,4.8\}$ & $\{50.86,50.99\}$ & $\{0.39,0.39\}$ & \cite{Becker13} \\
& $\{5.1\}$ & $\{51.00\}$ & $\{0.15\}$ & \cite{Becker21} \\
$Q_{\rm HII}$&&&&\\
& $\{7.0\}$ & $\{0.41\}$ & $\{0.13\}$ & \cite{Mason18} \\
& $\{7.6\}$ & $\{0.12\}$ & $\{0.07\}$ & \cite{Hoag19} \\
& $\{6.6,6.9,7.3\}$ & $\{0.30,0.50,0.55\}$ & $\{0.20,0.10,0.25\}$ & \cite{Konno14,Konno18} \\
& $\{7.6\}$ & $\{0.83\}$ & $\{0.10\}$ & \cite{Bolan22} \\
& $\{7.3\}$ & $\{0.49\}$ & $\{0.11\}$ & \cite{Greig22} \\
& $\{7.1,7.5\}$ & $\{0.48,0.60\}$ & $\{0.26,0.22\}$ & \cite{Davies18} \\
& $\{5.6,5.9\}$ & $\{0.04,0.06\}\;\;(\mathrm{up. lim.})$ & $\{0.05,0.05\}$ & \cite{McGreer15} \\
$\tau_{\rm es}$&&&&\\
& $\{\infty\}$ & $\{0.054\}$ & $\{0.007\}$ & \cite{Planck20} \\
\bottomrule
\end{tabularx}

\end{table}

We adopt flat priors $\pi(\theta)$ on the parameters within the ranges $f_{\rm esc}\in [0,1]$, $M_{\rm UV}^{\rm lim}\in [-20,-8]$, and $X\in [0,15]$. Note that the latter is the same for any DM scenario; it refers to different units depending on the meaning of $X$; e.g., in the WDM scenario, $X$ represents the DM particle mass $m_X$ in units of keVs. We then sample the posterior distribution $\mathcal{P}(\theta)\propto \mathcal{L}(\theta)\, \pi(\theta)$ by running $\texttt{emcee}$ with $10^5$ steps and $300$ walkers; each walker is initialized with a random position uniformly sampled from the (flat) priors.
After checking the auto-correlation time, we remove the first $20\%$ of the flattened chain to ensure the burn-in; the typical acceptance fractions of the various runs are in the range 30--40\%. Convergence of the chains and decently shaped posteriors are attained in a reasonable computational time, around 90--120 min on a laptop with an \texttt{Intel i7-8565U} CPU running the MCMC algorithm parallelized over $8$ cores.

\section{Results and Discussion}\label{sec|results}

\begin{figure}[H]
\includegraphics[width=0.46\textwidth]{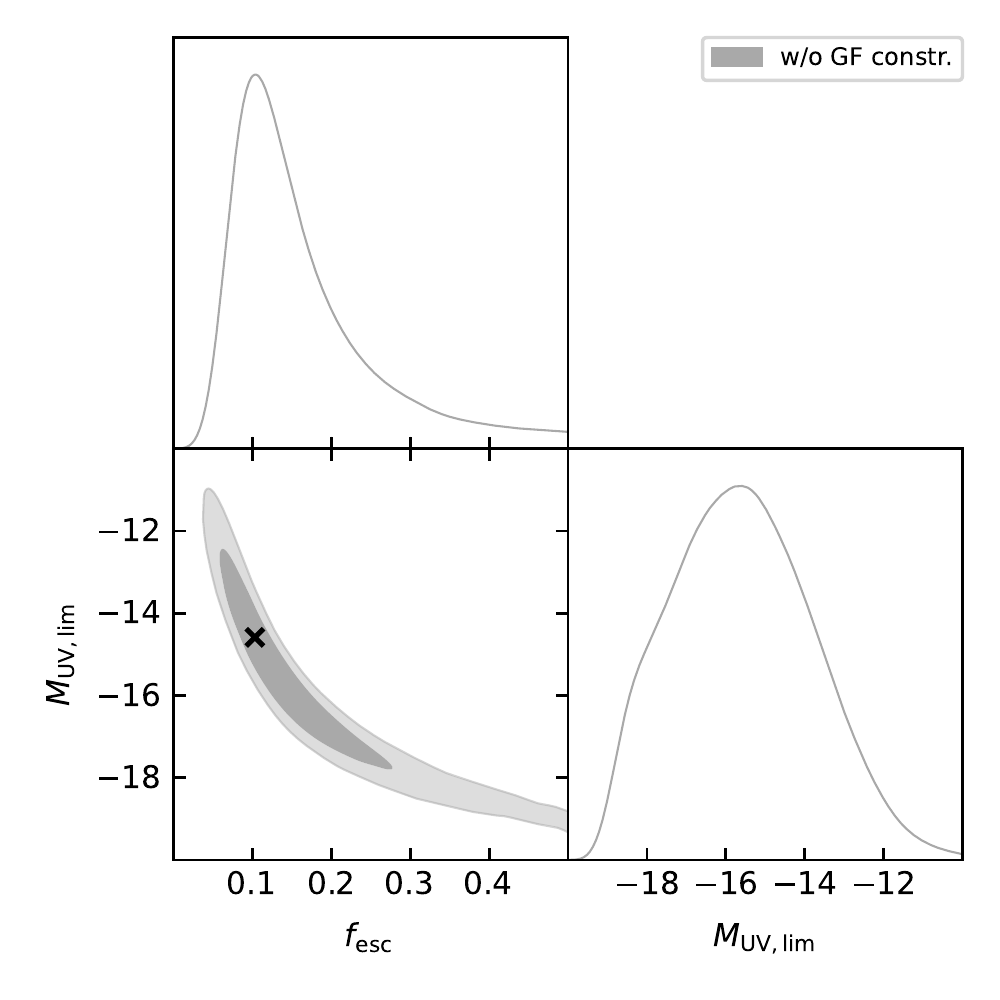}
\includegraphics[width=0.48\textwidth]{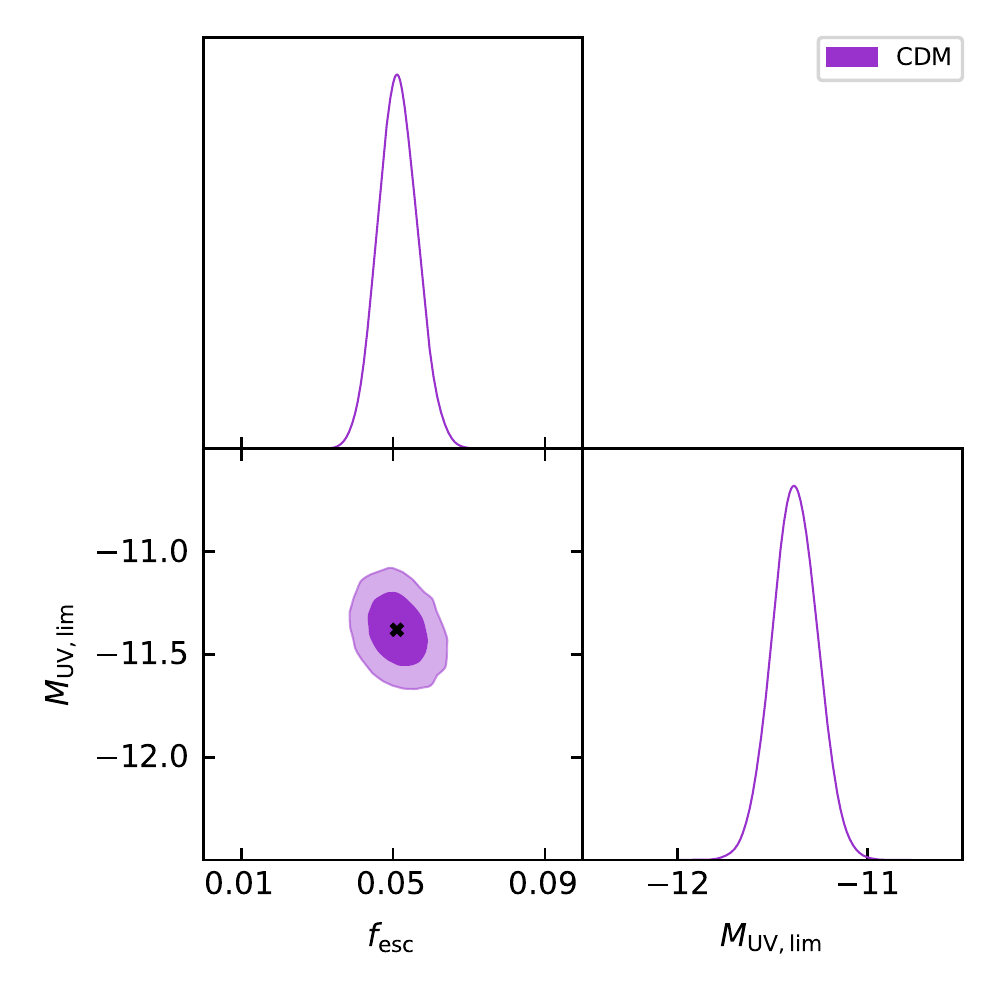}
\caption{MCMC posterior distributions in the standard CDM scenario for the escape fraction $f_{\rm esc}$ of ionizing photons and the limiting UV magnitude $M_{\rm UV,lim}$.
In the left panel (grey contours/lines), the galaxy formation constraint derived via abundance matching is not included in the likelihood (see Section \ref{sec|Bayes}), while in the right panel (magenta contours/lines), this is taken into account.
The contours show $68\%$ and $95\%$ confidence intervals, the black cross shows the maximum likelihood position, and the marginalized distributions are in arbitrary units (normalized to $1$ at their \mbox{maximum value)}.}\label{fig|CDM}
\end{figure}

In this section, we report and discuss our results for different DM scenarios, according to the Bayesian procedure presented in Section \ref{sec|Bayes}. As a preliminary step, we analyze the data without taking into account the galaxy formation constraint from abundance matching. The result is shown by the grey contours/lines in Figure~\ref{fig|CDM} (left panel). The marginalized constraint on the escape fraction and the limiting UV magnitude read \mbox{$f_{\rm esc}\approx 0.16^{+0.03}_{-0.01}$} and $M_{\rm UV}^{\rm lim}\approx -15.6^{+1.7}_{-2.0}$. As expected these are rather loose, since there is a clear degeneracy between these two quantities: the data on reionization history can be reproduced in principle by assuming a smaller $f_{\rm esc}$, i.e., decreasing the number of ionizing photons escaping from each galaxy, while increasing the limiting UV magnitude $M_{\rm UV}^{\rm lim}$, i.e., enhancing the number of galaxies that populate the faint end of the luminosity function and, hence, contribute to the ionizing background.

Once the galaxy formation constraint $M_{\rm H}(M_{\rm UV}^{\rm lim},z|X) = M_{\rm H}^{\rm GF}$ from abundance matching is included in the statistical analysis, the result becomes sensitive to the DM scenario. We illustrate the outcome for the standard CDM by the magenta contours/lines in Figure~\ref{fig|CDM} (right panel). In this case, the DM particle is so cold ($\gtrsim$GeV) that independently of its precise value the threshold halo mass for galaxy formation $M_{\rm H}^{\rm GF}$ corresponds uniquely
to a faint $M_{\rm UV}^{\rm lim}$ and correspondingly to a quite low $f_{\rm esc}$, see the solid black line in Figure~\ref{fig|abma}. The marginalized constraints for CDM turns out to be $f_{\rm esc}\approx 0.051^{+0.005}_{-0.005}$ and $M_{\rm UV}^{\rm lim}\approx -11.4^{+0.1}_{-0.1}$.

\begin{figure}[H]
\includegraphics[width=0.8\textwidth]{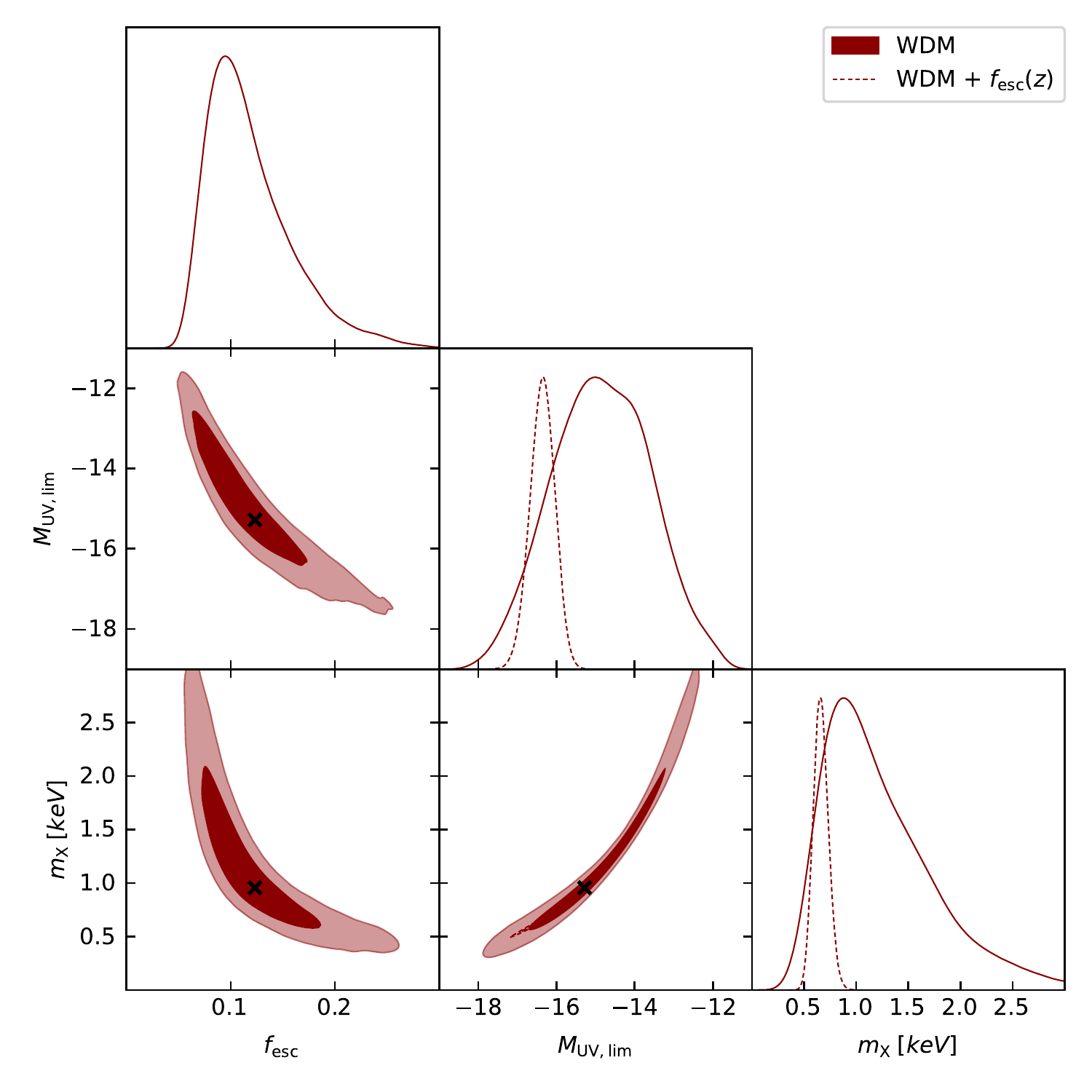}
\caption{MCMC posterior distributions in the WDM scenario (red contours/lines), for the escape fraction $f_{\rm esc}$ of ionizing photons, the limiting UV magnitude $M_{\rm UV,lim}$, and the DM particle's mass $m_X$.
The dashed lines refer to a run where the escape fraction has been set to the
redshift dependent values from the radiative transfer simulations
{by} \cite{Puchwein19}. The contours show $68\%$ and $95\%$ confidence intervals, the black cross shows the maximum likelihood position, and the marginalized distributions are in arbitrary units (normalized to $1$ at their maximum value).}\label{fig|WDM}
\end{figure}

In the other DM scenarios, the situation is drastically different, because the
galaxy formation constraint from abundance matching depends crucially on the DM astroparticle property $X$. The results for WDM are illustrated by the red lines/contours in Figure~\ref{fig|WDM}. For such a case, it is seen from Figure~\ref{fig|abma} that the critical halo mass for galaxy formation $M_{\rm H}^{\rm GF}$ corresponds to a UV-limiting magnitude $M_{\rm UV}^{\rm lim}\approx -15$, appreciably brighter with respect to CDM for particle masses $m_X\sim$ keV, while it converges to the CDM value $\approx -11.5$ for $m_X\gtrsim$ some keVs. However, the MCMC algorithm, informed by the data on cosmic reionization, disfavor solutions with high values of $m_X$ that will correspond to very faint limiting UV magnitude and small escape fraction, with respect to values of $m_X\approx 1$ keV, which will instead maximize the likelihood in the subspace of the astrophysical parameters $f_{\rm esc}$ and $M_{\rm UV}^{\rm lim}$ (see black crosses in the posterior contours). In all, for WDM, we find the marginalized constraints to read
$f_{\rm esc}\approx 0.12^{+0.02}_{-0.05}$,
$M_{\rm UV}^{\rm lim}\approx -14.8^{+1.2}_{-1.2}$ and $m_X\approx 1.3^{+0.3}_{-0.7}$ keV.

\begin{figure}[H]

\includegraphics[width=0.8\textwidth]{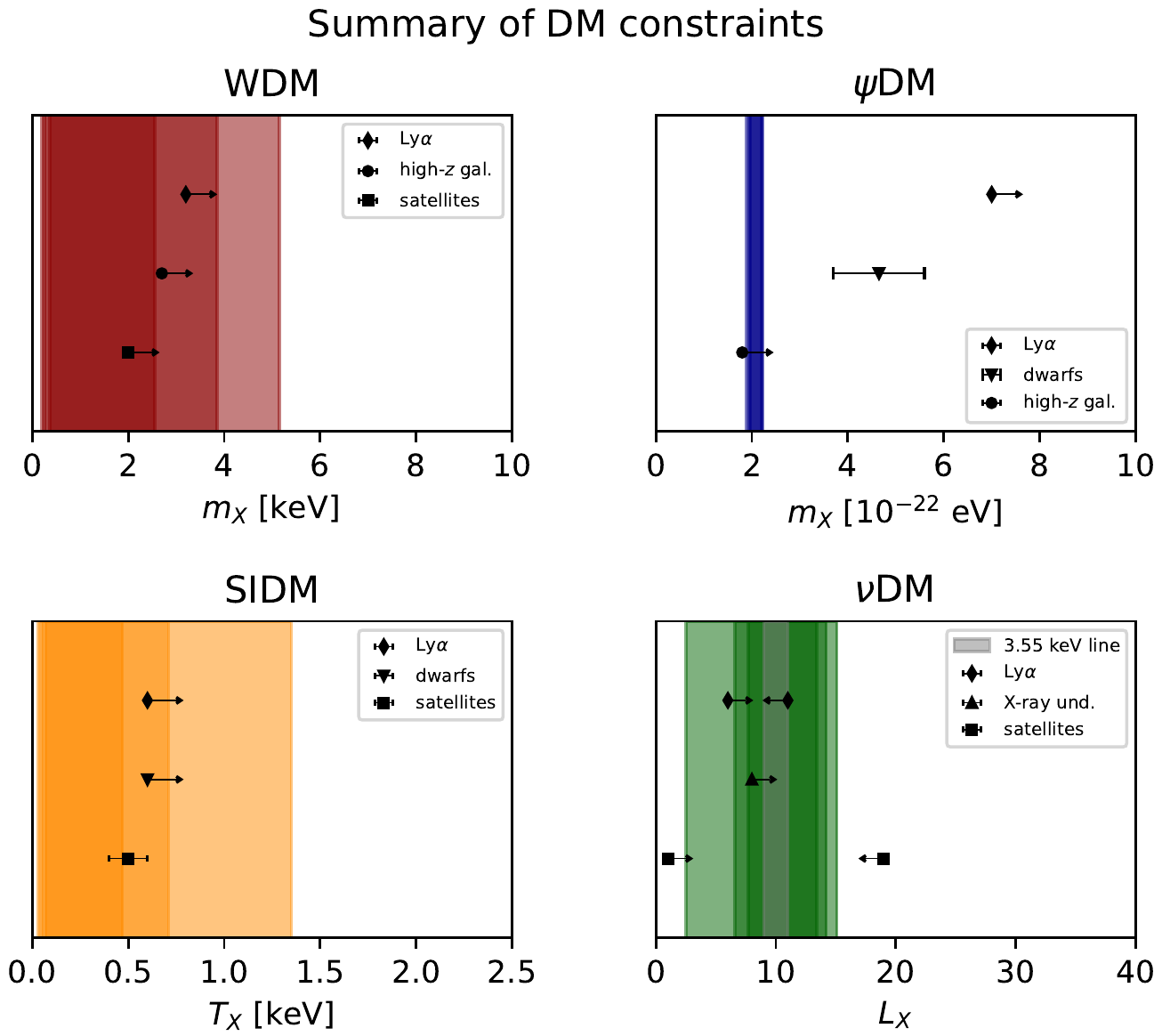}
\caption{Summary of the astroparticle constraints in different DM scenarios. The colored shaded areas illustrate the constraints from this work for confidence intervals of $2\sigma$, $3\sigma$, and $5\sigma$ (from darker to lighter shades). The black points show the literature constraints from independent observations: for WDM (top left panel) from \cite{Irsic17wdm} (diamond), \cite{Shirasaki21} (circle) and \cite{Newton21} (square); for $\psi$DM (top right panel) from \cite{Irsic17fdm} (diamond and circle) and \cite{Calabrese16} (inverted triangle); for SIDM (bottom left panel) \mbox{from \cite{Irsic17wdm}} (diamond), \cite{Vogelsberger16} (inverted triangle) and \cite{Huo18} (square); for $\nu$DM (bottom right panel) \mbox{from \cite{Vegetti18}} (grey shaded area), \cite{Irsic17wdm} (diamond), \cite{Horiuchi14} (triangle) and \cite{Lovell16} (square).}\label{fig|APP}
\end{figure}

The added value of our estimate for $m_X$ appears evident when considered in comparison and/or combination with independent data from, e.g., the Lyman-alpha \mbox{forest \cite{Viel13,Irsic17wdm}}, high-redshift galaxy counts \cite{Menci16,Shirasaki21}, integrated 21cm emission \cite{Carucci15,Boyarsky19,Chatterjee19}, and Milky Way satellite counts \cite{Kennedy14,Newton21}, as illustrated in Figure~\ref{fig|APP}.
These classic probes provide lower bounds $m_X\gtrsim$ 2--3 keVs at $2\sigma$, which are only marginally consistent with the tail of our posterior distribution. Note also that other numerical studies have shown that, to obtain the observed kpc cores of dwarf galaxies, a thermal relic mass as low as $m_X\sim 0.1$ keV may be \mbox{needed \cite{Maccio12}}, which is certainly excluded by our analysis and by the other probes listed above. In all, the tensions among all these independent constraints tend to lower the case for the keV-scale thermal WDM as a viable alternative to CDM.

In Figure~\ref{fig|WDM}, we also illustrate the results (dashed lines) when the escape fraction is set to the redshift-dependent value $f_{\rm esc}(z)\approx \min[6.9\times 10^{-5}\,(1+z)^{3.97},0.18]$, which increases from $\approx$0.05 in the local Universe up to $\approx$0.2 at high redshift. This parameterization is based on the radiative transfer simulations by \cite{Puchwein19}, which have been gauged to reproduce the evolution of the overall ionizing photon rate with cosmic time.  At the redshifts relevant for reionization $f_{\rm esc},\lesssim 0.2$ appreciably exceeds the value obtained in the run with free $f_{\rm esc}$, a smaller number of galaxies is needed to meet the reionization constraints. As a consequence, we estimate a brighter value of $M_{\rm UV}^{\rm lim}\approx
-16.4^{+0.3}_{-0.3}$ and correspondingly a smaller
$m_X\approx 0.66^{+0.07}_{-0.08}$ keV.

\begin{figure}[H]

\includegraphics[width=0.8\textwidth]{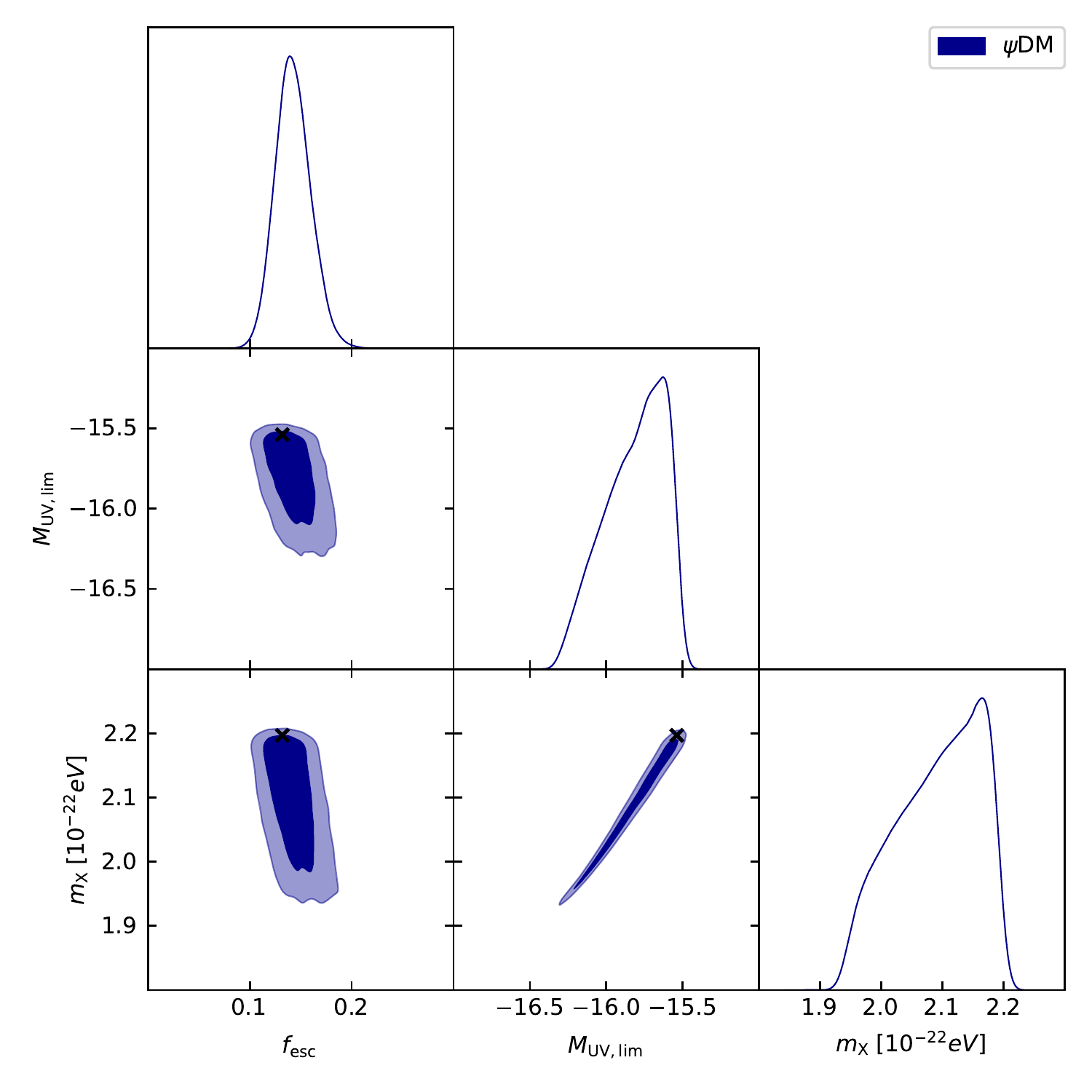}
\caption{MCMC posterior distributions in the $\psi$DM scenario (blue contours/lines), for the escape fraction $f_{\rm esc}$ of ionizing photons, the limiting UV magnitude $M_{\rm UV,lim}$, and particle mass $m_X$. The contours show  $68\%$ and $95\%$ confidence intervals, the black cross shows the maximum likelihood position, and the marginalized distributions are in arbitrary units (normalized to $1$ at their \mbox{maximum value)}.}\label{fig|psiDM}
\end{figure}

The situation in other DM scenarios is somewhat similar to WDM. The main difference resides in the behavior of the halo mass function (at small masses), which induces a different shape of the relationship between $M_{\rm H}$ and $M_{\rm UV}$, and in turn, this affects the galaxy formation constraint. In the $\psi$DM case, whose results are illustrated in Figure~\ref{fig|psiDM}, the $M_{\rm H}(M_{\rm UV},z|m_X)$ relation is quite steep since the halo mass function bends down for small masses. This implies that
the galaxy formation constraint is tight, because a small variation in the
particle mass $m_X$ may induce $M_{\rm H}^{\rm GF}$ to correspond to
appreciably different $M_{\rm UV}^{\lim}$. As a consequence,
the marginalized distribution of $m_X$ is extremely narrow.
Our posterior estimates for $\psi$DM are found to be
$f_{\rm esc}\approx 0.14^{+0.02}_{-0.02}$,
$M_{\rm UV}^{\rm lim}\approx -15.8^{+0.3}_{-0.1}$ and $m_X\approx 2.09^{+0.09}_{-0.05}\times 10^{-22}$ eV.
The latter value is appreciably smaller than independent constraints present in the literature (see Figure~\ref{fig|APP}), which are somewhat in tension among themselves, from high-redshift galaxy counts \cite{Schive16,Kulkarni22}, ultra-faint dwarfs \cite{Calabrese16}, and Ly$\alpha$ forest \cite{Irsic17fdm}. Note also that galaxy scaling relations highlight the difficulties of $\psi$DM in solving the small-scale problems of CDM \cite{Burkert20}. Currently the evidence for $\psi$DM as a viable alternative to CDM is marginal and should be reconsidered in light of future data.

\begin{figure}[H]

\includegraphics[width=0.8\textwidth]{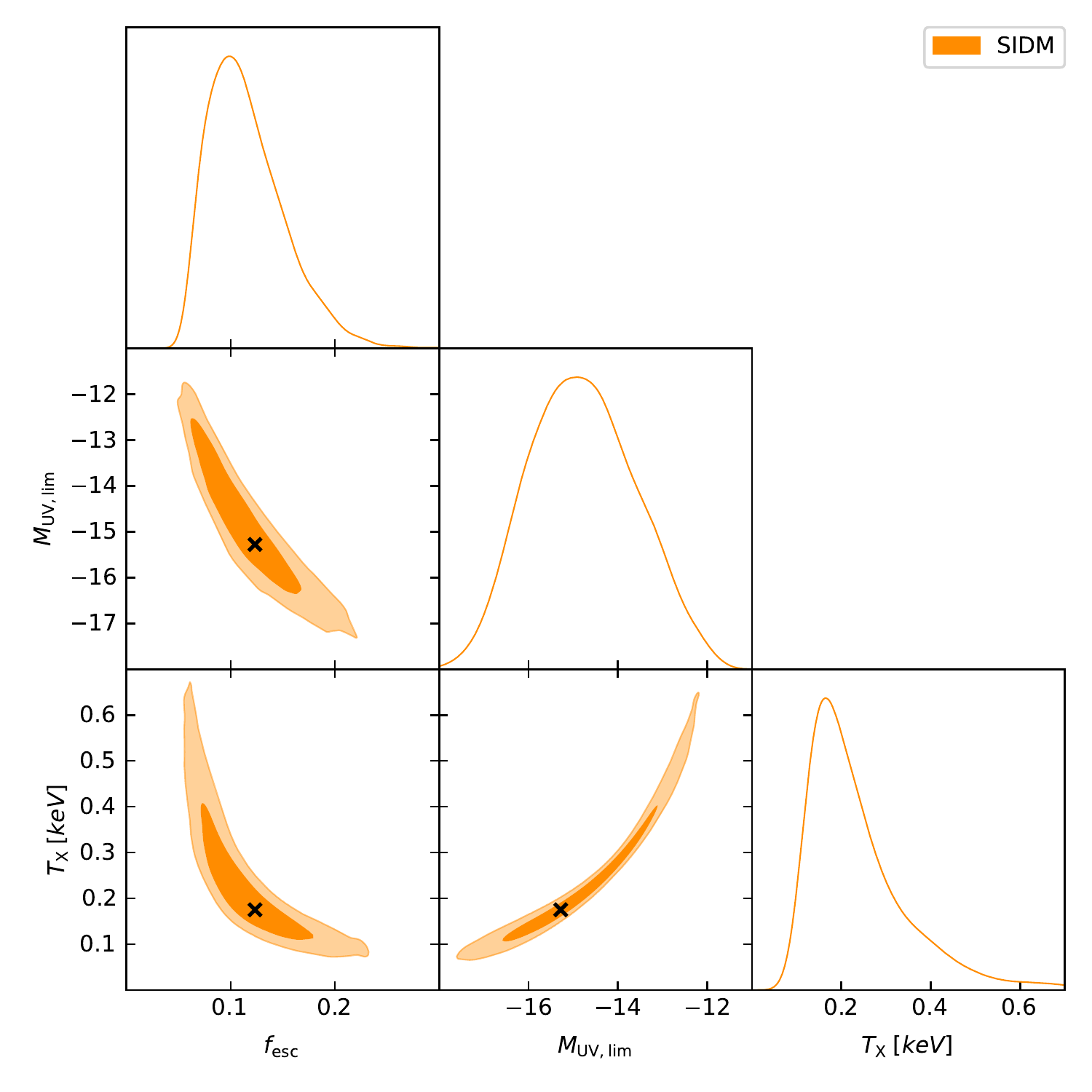}
\caption{MCMC posterior distributions in the SIDM scenario (orange contours/lines), for the escape fraction $f_{\rm esc}$ of ionizing photons, the limiting UV magnitude $M_{\rm UV,lim}$, and visible sector temperature $T_X$ at kinetic decoupling. The contours show $68\%$ and $95\%$ confidence intervals, the black cross shows the maximum likelihood position, and the marginalized distributions are in arbitrary units (normalized to $1$ at their maximum value).}\label{fig|SIDM}
\end{figure}

In the SIDM scenario, whose results are shown in Figure~\ref{fig|SIDM},
the abundance matching relation $M_{\rm H}(M_{\rm UV}^{\rm lim},z|T_X)$ changes quite abruptly for small values of the temperature $T_X$ at kinetic decoupling. In particular, temperatures $T_X\lesssim 0.2$ keV correspond to limiting UV magnitudes brighter than $-17$, which are excluded by the present data on the UV luminosity function that steeply goes down to such values;
this is why the posterior
of $T_X$ is somewhat truncated toward low values. The marginalized estimates for SIDM are $f_{\rm esc}\approx 0.12^{+0.02}_{-0.05}$,
$M_{\rm UV}^{\rm lim}\approx -14.8^{+1.1}_{-1.4}$ and $T_X\approx 0.24^{+0.04}_{-0.13}$ keV. As shown in Figure~\ref{fig|APP}, the latter value is consistent within $3\sigma$ with that from Ly$\alpha$ forest \cite{Irsic17wdm}, satellite counts \cite{Huo18} and dwarf galaxies \cite{Vogelsberger16}.

\begin{figure}[H]

\includegraphics[width=0.8\textwidth]{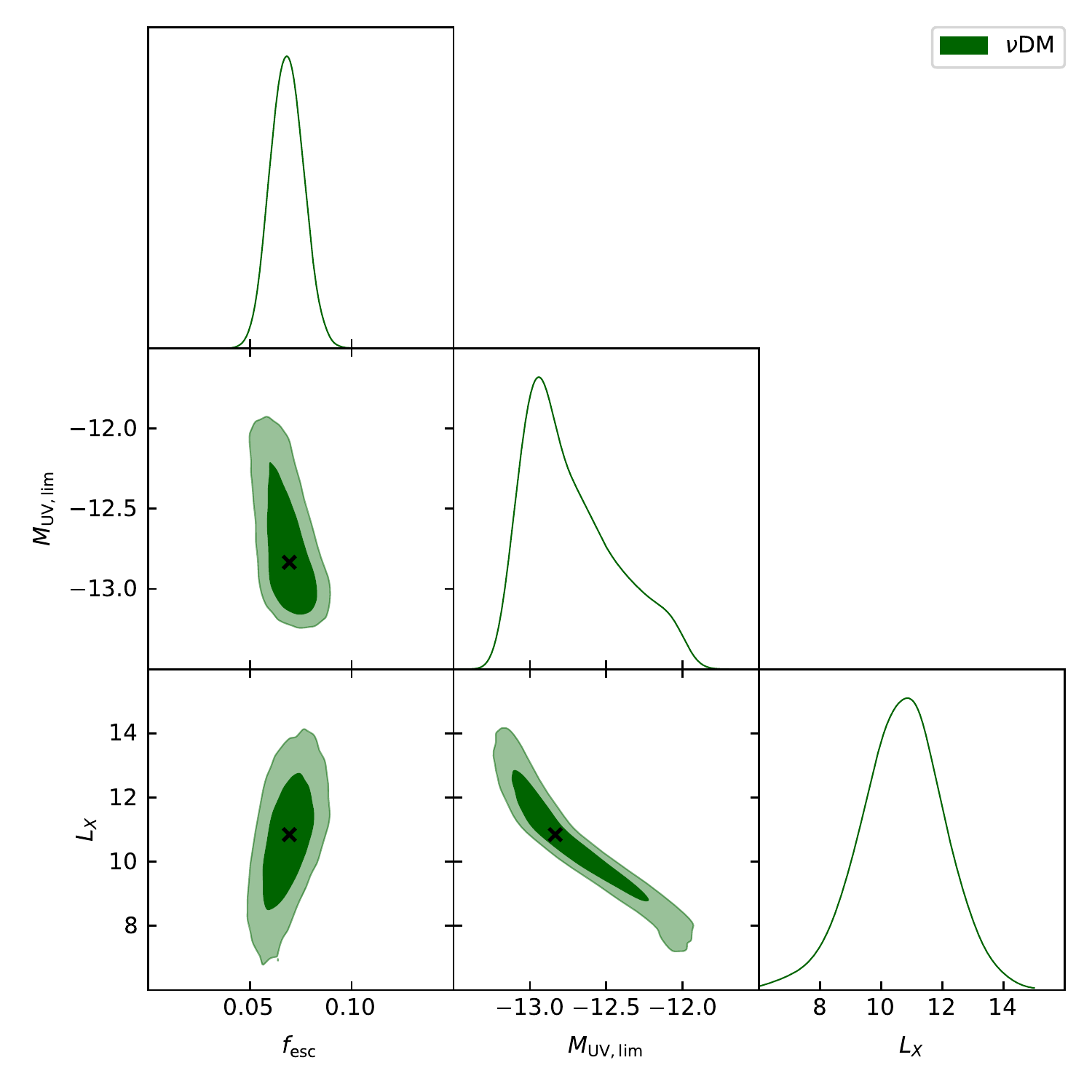}
\caption{MCMC posterior distributions in the $\nu$DM scenario (sterile neutrino DM with a mass of $7$~keV), for the escape fraction $f_{\rm esc}$ of ionizing photons, the limiting UV magnitude $M_{\rm UV,lim}$, and the lepton asymmetry parameter $L_X$. The contours show $68\%$ and $95\%$ confidence intervals, the black cross shows the maximum likelihood position, and the marginalized distributions are in arbitrary units (normalized to $1$ at their maximum value).}\label{fig|nuDM}
\end{figure}

In the $\nu$DM scenario with a particle mass $m_X\sim 7$ keVs, whose results are shown in Figure~\ref{fig|nuDM}, the derived constraints are less sharp. This is because, as it can be seen from Figure~\ref{fig|abma},  the $M_{\rm H}(M_{\rm UV},z|L_X)$ relation from abundance matching changes only slightly for different lepton asymmetries $L_X$, implying a quite broad posterior on such parameter. Specifically, for $\nu$DM, we find marginalized estimates amounting to $f_{\rm esc}\approx 0.068^{+0.008}_{-0.008}$,
$M_{\rm UV}^{\rm lim}\approx -12.7^{+0.2}_{-0.4}$ and $L_X\approx 10.7^{+1.4}_{-1.4}$. These constraints for 7 keV $\nu$DM are consistent with those present in the literature (see Figure~\ref{fig|APP}), satellite counts in the Milky Way \cite{Lovell16}, gravitational lensing observations \cite{Vegetti18}, X-ray non-detections \cite{Horiuchi14}, and Ly$\alpha$ \mbox{measurements \cite{Irsic17wdm}}, which concurrently place the lepton asymmetry in the range of $L_X\sim 8$--12.

\begin{figure}[H]

\includegraphics[width=11cm]{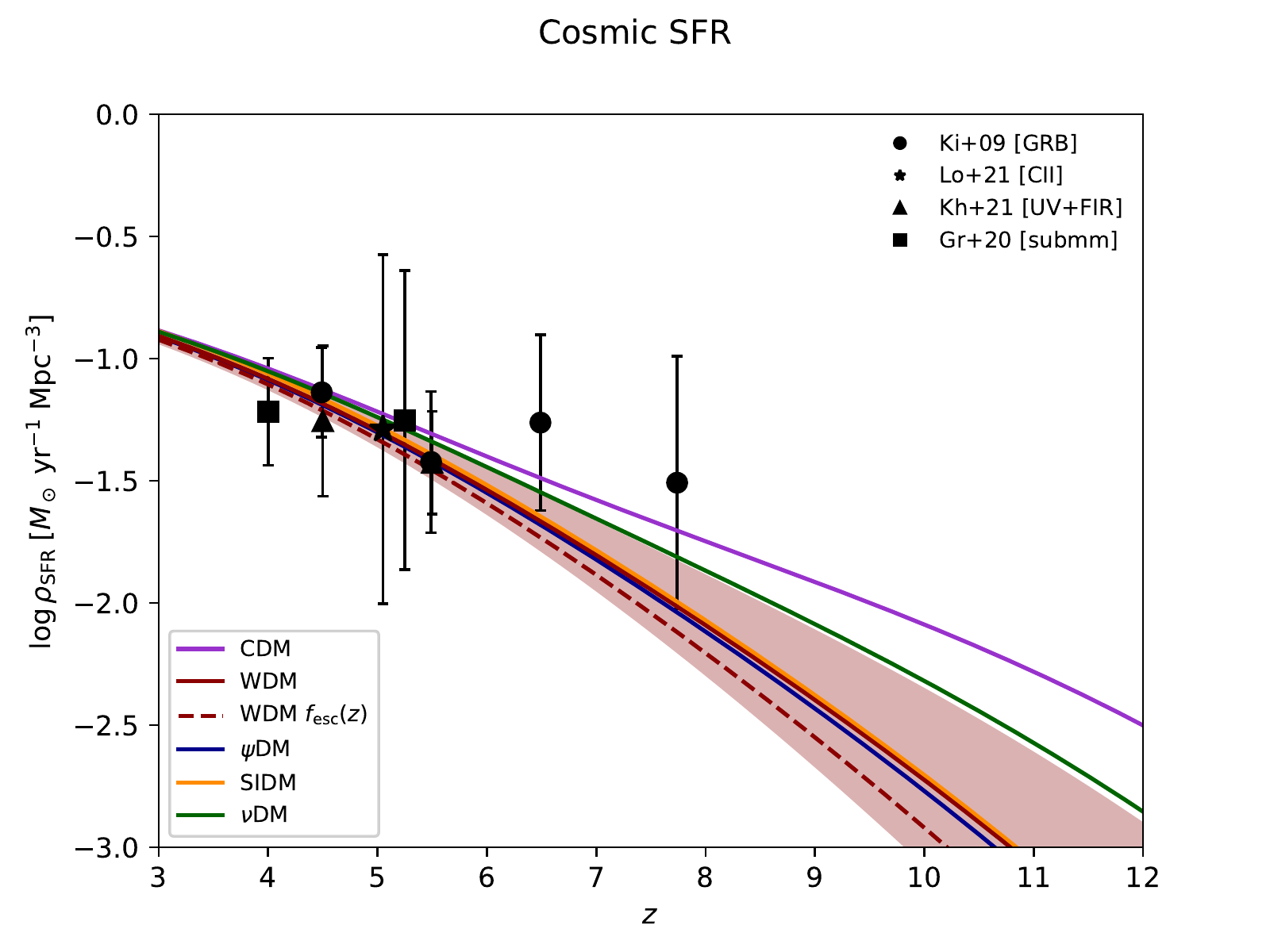}
\caption{The cosmic SFR density as a function of redshift. Data are from GRBs (circles; {see} \cite{Kistler09}), CII (stars; see \cite{Loiacono21}), UV+FIR (triangles; see \cite{Khusanova21}), and (sub)mm (squares; see \cite{Gruppioni20}). Lines illustrate the best fits from the MCMC analysis in various DM scenarios:  CDM (purple solid), WDM (red solid), WDM with a redshift-dependent $f_{\rm esc}(z)$ (red dashed), $\psi$DM (blue solid), SIDM (orange solid) and $\nu$DM (green solid). The typical $2\sigma$ credible interval from sampling the posterior distribution is shown, for clarity, only in the WDM scenario, as a red shaded area. Note that the cosmic SFR density has not been fitted upon, since the related measurements have not been exploited in constructing the likelihood of our Bayesian analysis.}\label{fig|SFR}
\end{figure}

\begin{figure}[H]

\includegraphics[width=11cm]{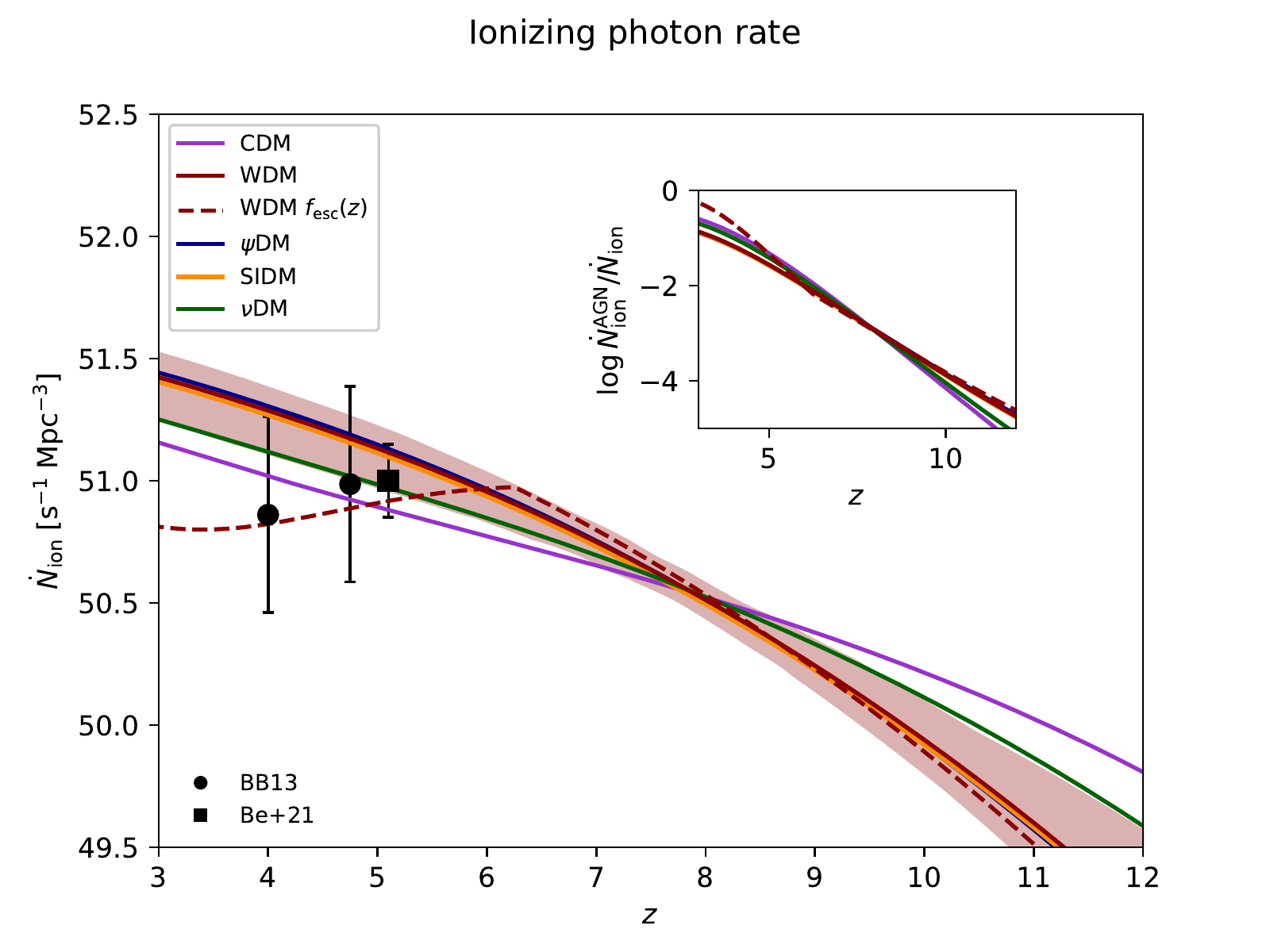}
\caption{The ionizing photon rate as a function of redshift. Data are {from} \cite{Becker13} (circles) and \cite{Becker21} (squares). Lines as in Figure~\ref{fig|SFR}. The inset illustrates the contribution of AGNs (see Equation \eqref{eq|Ndotion_AGN}) to the total ionizing photon rate as a function of redshift, in the various DM scenarios.}\label{fig|Ndotion}
\end{figure}

\begin{figure}[H]

\includegraphics[width=11cm]{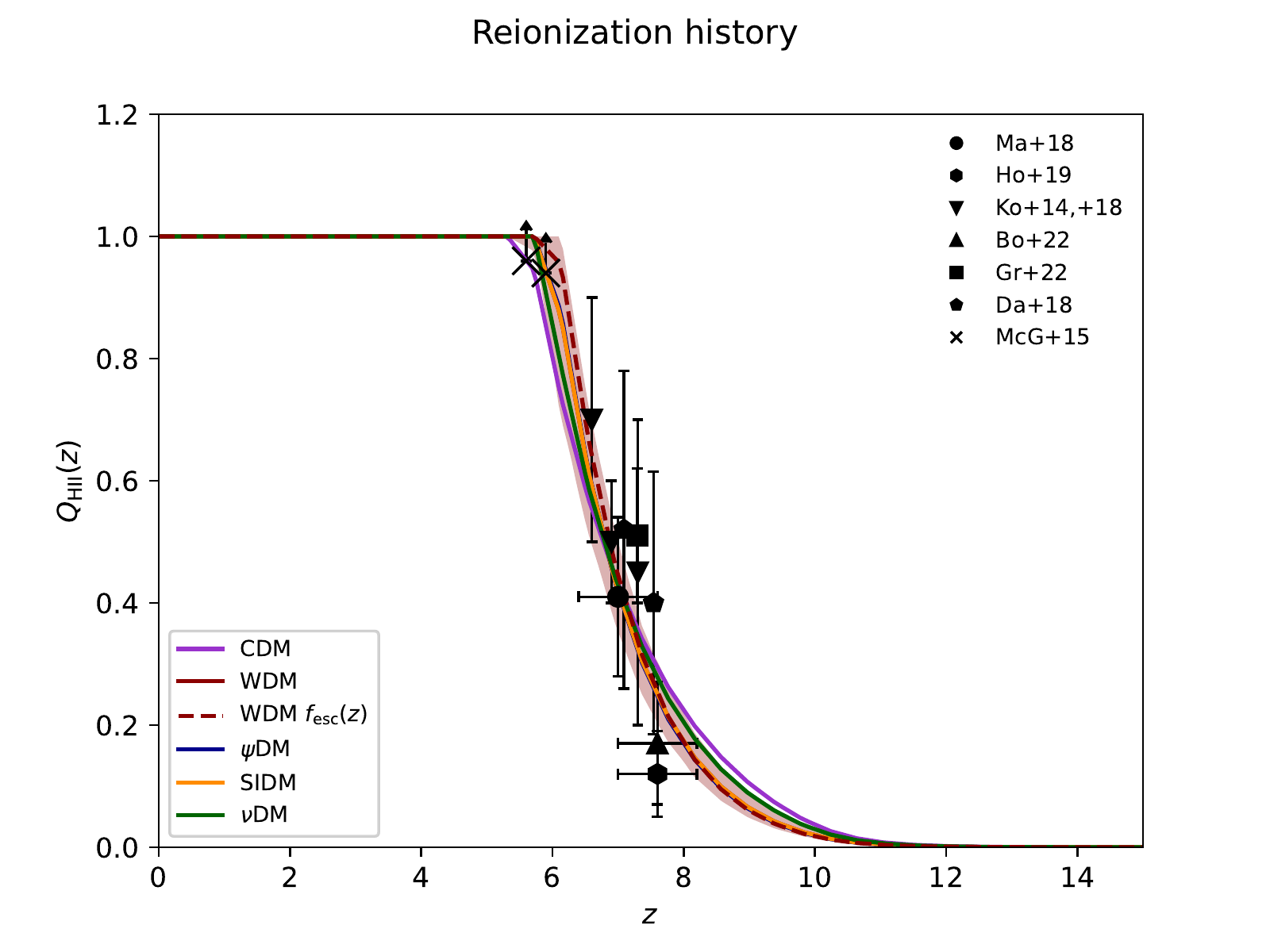}
\caption{The reionization history of the Universe, in terms of the volume filling fraction $Q_{\rm HII}$ of ionized hydrogen as a function of redshift $z$. Data are {from} \cite{Mason18} (circle), \cite{Hoag19} (hexagon), \cite{Konno14,Konno18} (inverted triangle), \cite{Bolan22} (triangle), \cite{Greig22} (squares), \cite{Davies18} (pentagon), and \cite{McGreer15} (crossed). Lines as in Figure~\ref{fig|SFR}.}\label{fig|QHII}
\end{figure}

\begin{figure}[H]
\includegraphics[width=12cm]{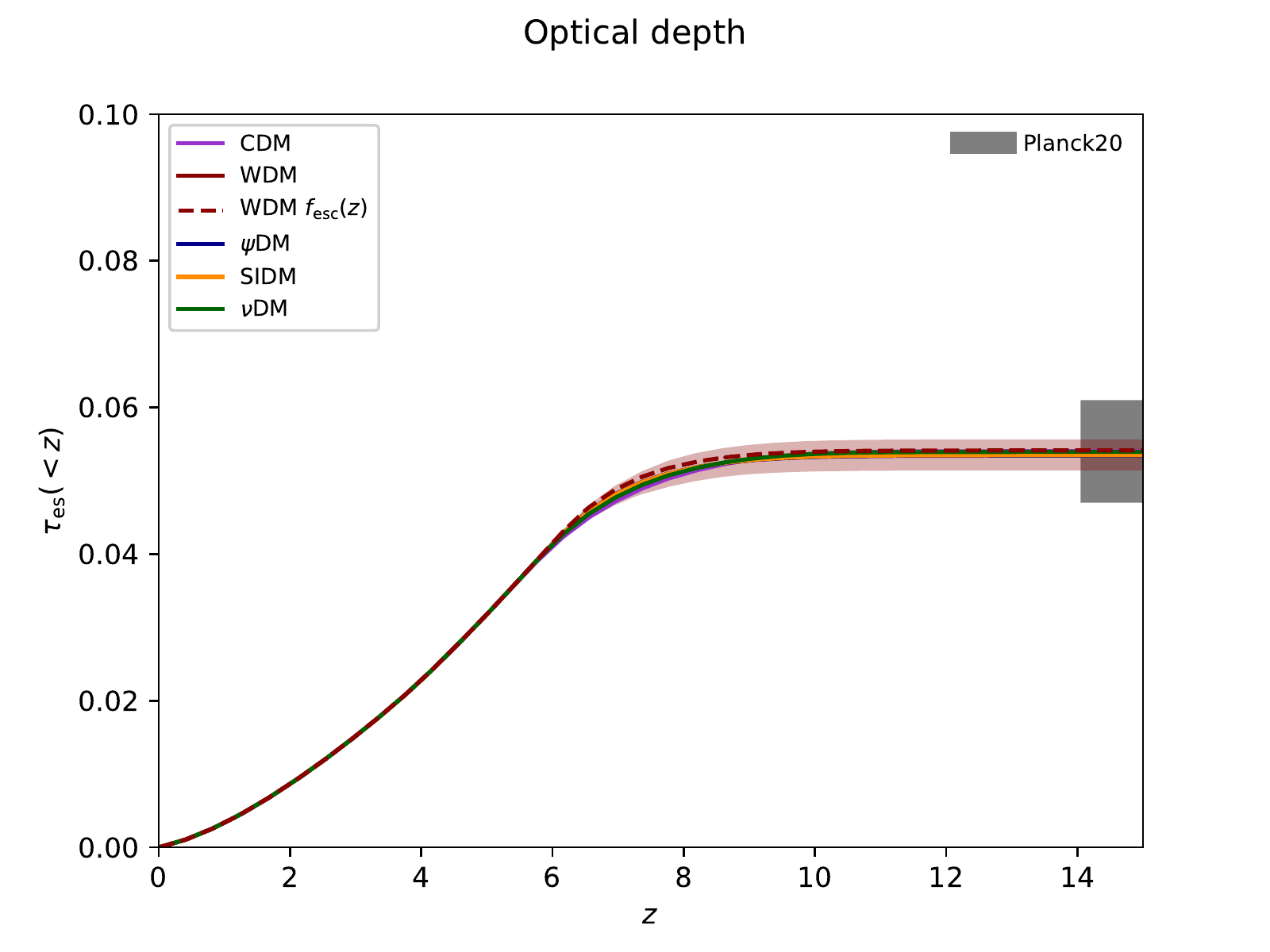}
\caption{The optical depth to electron scattering $\tau_{\rm es}(<z)$ as a function of redshift $z$. Data are {from}~\cite{Planck20} (shaded area). Lines as in Figure~\ref{fig|SFR}.}\label{fig|tau_es}
\end{figure}

The marginalized posterior estimates in the different DM scenarios are summarized in Table \ref{tab|results}. Moreover, we show in Figures \ref{fig|SFR}--\ref{fig|tau_es} the behavior of our best-fit models on various observables, which include the cosmic SFR density, the ionizing photon rate (the contribution of AGNs, illustrated in the inset, is shown to be minor for $z\gtrsim 6$ and subdominant at lower redshift), the volume filling factor of ionized hydrogen and the electron scattering optical depth. Note that the cosmic SFR rate density has not been exploited in building up the likelihood in Section \ref{sec|Bayes} and thus it has not been fitted upon. In fact, our best-fit models refer to the cosmic SFR rate density when the UV luminosity function is integrated down to a magnitude $M_{\rm UV}^{\rm lim}$, which may be appreciably fainter than the values corresponding to the observational determinations (e.g., the ones based on current UV data refer to a limit $M_{\rm UV}\approx -17$). Nevertheless, it is remarkable that our best-fit models turn out to be consistent with the available observational constraints on the cosmic SFR density from GRBs by \cite{Kistler09}, CII by \cite{Loiacono21}, UV+FIR by \cite{Khusanova21}, and (sub)mm by \cite{Gruppioni20}, at least for $z\lesssim 8$.
On the other hand, we stress that the predictions on the cosmic SFR density for the non-standard DM scenarios tend to significantly deviate with respect to that of CDM as soon as the redshift increases much beyond $z\gtrsim 8$. Therefore, upcoming precision determinations of the cosmic SFR at $z\gtrsim 10$, as recently attempted with early JWST data \mbox{by \cite{Harikane22}}, could provide relevant additional constraints on the DM scenario.

\vspace{-9pt}

\begin{table}[H]
\caption{Marginalized posterior estimates of the parameters from the MCMC analysis for the different DM scenarios considered in the main text. Specifically, $f_{\rm esc}$ is the escape fraction, $M_{\rm UV}^{\rm lim}$ is the limiting UV magnitude, and the astroparticle quantity $X$ in the third column stands for: $m_X$ is in units of keV for WDM scenario, $m_X$ in units of $10^{-22}$ eV for the $\psi$DM scenario, $T_X$ in units of keV for the SIDM scenario, and $L_X$ for the $\nu$DM scenario. Mean and $1\sigma$ confidence limits are reported. The last column refers to the value of the Bayes information criterion (BIC) for model comparison, see Section \ref{sec|results}.}\label{tab|results}
\newcolumntype{C}{>{\centering\arraybackslash}X}
\begin{tabularx}{\textwidth}{CCCCC}
\toprule

\textbf{Scenario} &\boldmath{\textbf{$f_{\rm esc}$} }&\boldmath{ \textbf{$M_{\rm UV, lim}$}} & \boldmath{\textbf{$X$}} & \textbf{BIC} \\
\midrule
w/o GF  & $0.16^{+0.03}_{-0.01}$ &  $-15.6^{+1.7}_{-2.0}$ & $-$ & $-$\\
CDM & $0.051^{+0.005}_{-0.005}$ &   $-11.4^{+0.1}_{-0.1}$ & $-$ & $53.8$\\
WDM & $0.12^{+0.02}_{-0.05}$ & $-14.8^{+1.2}_{-1.2}$ & $1.3^{+0.3}_{-0.7}$ & $36.7$ \\
WDM & $f_{\rm esc}(z)$ & $-16.4^{+0.3}_{-0.3}$ & $0.66^{+0.07}_{-0.08}$ & $-$ \\
$\psi$DM &  $0.14^{+0.02}_{-0.02}$ &  $-15.8^{+0.3}_{-0.1}$ & $2.09^{+0.09}_{-0.05}$ & $37.1$\\
SIDM &  $0.12^{+0.02}_{-0.05}$ &  $-14.8^{+1.1}_{-1.4}$ & $0.24^{+0.04}_{-0.13}$ & $36.9$\\
$\nu$DM &  $0.068^{+0.008}_{-0.008}$ &  $-12.7^{+0.2}_{-0.4}$ & $10.7^{+1.4}_{-1.4}$ & $39.6$\\
\bottomrule
\end{tabularx}

\end{table}

\vspace{-9pt}

\vspace{-10pt}

As for the reionization observables $\dot N_{\rm ion}$, $Q_{\rm HII}$ and $\tau_{\rm es}$, it is worth stressing that all our best-fit models perform comparably well in reproducing the available data. This is also highlighted by the $2\sigma$ credible interval from sampling the posterior distribution, which is shown only in the WDM case for clarity (red shaded area). In terms of projection on these observables, the different DM scenarios are consistent with each other, approximately within $2\sigma$, while they differ appreciably from the standard CDM case. In the same vein, we can also attempt a model comparison via the Bayes information criterion \cite{Liddle04} defined as BIC$\equiv -2\, \ln\mathcal{L_{\rm max}}+N_{\rm par}\, \ln N_{\rm data}$ in terms of the maximum likelihood estimate $\mathcal{L_{\rm max}}$, of the number of parameters $N_{\rm par}$, and the number of data points $N_{\rm data}$. The BIC comes from approximating the Bayes factor, which gives the posterior odds of one model against another, presuming that the models are equally favored a priori. Note that what matters is the relative value of the BIC among different models; in particular, a difference of around ten or more indicates evidence in favor of the model with the smaller value. The values of the BIC (for the different DM scenarios) are reported in Table \ref{tab|results}. Taken at face value, the BIC suggests evidence in favor of the scenarios alternative to CDM, though it is risky to recognize a preference among them.

\begin{figure}[H]
\begin{tabular}{@{}ccc}
\includegraphics[width=4.5cm,height=4.5cm]{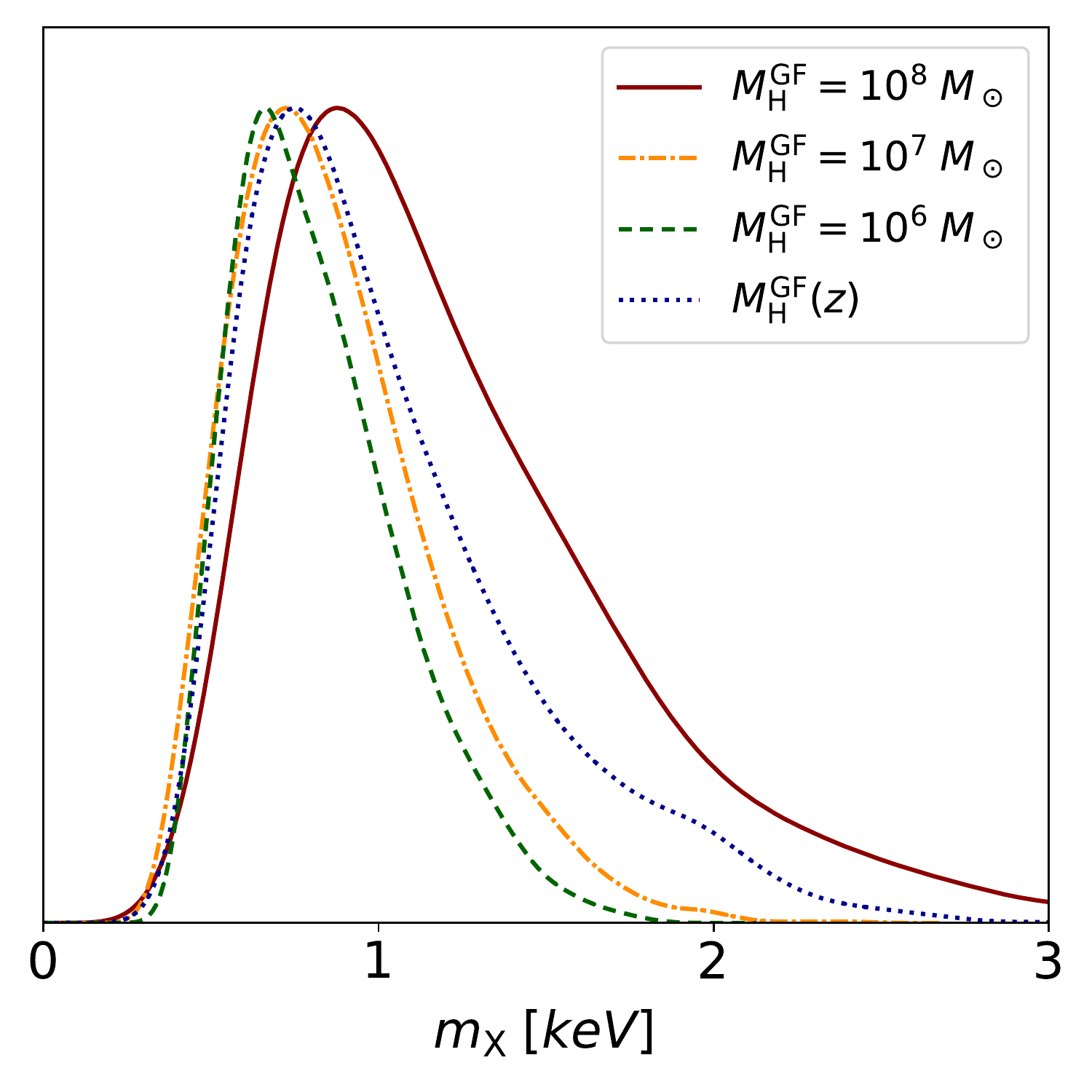}
\includegraphics[width=4.5cm,height=4.5cm]{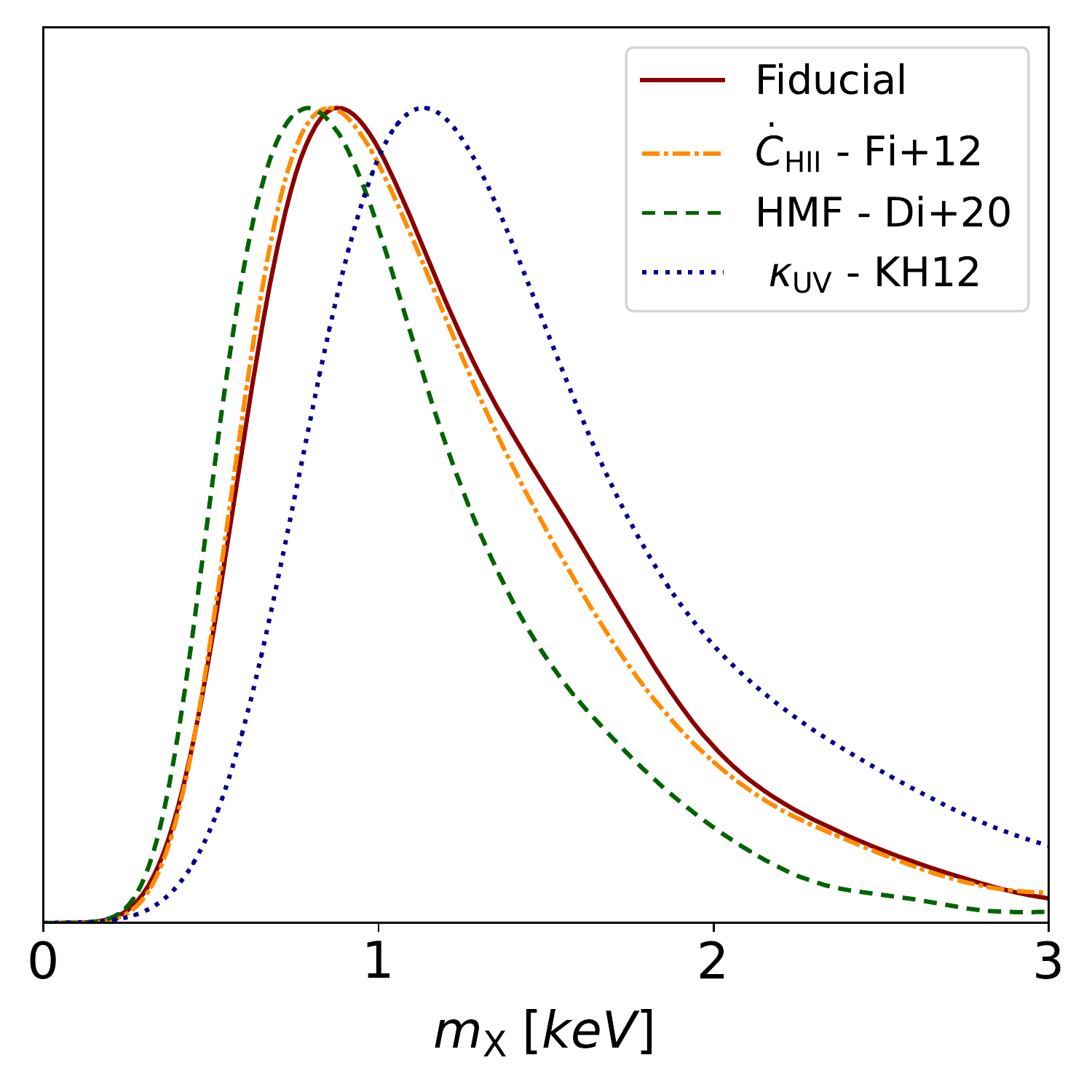}
\includegraphics[width=4.5cm,height=4.5cm]{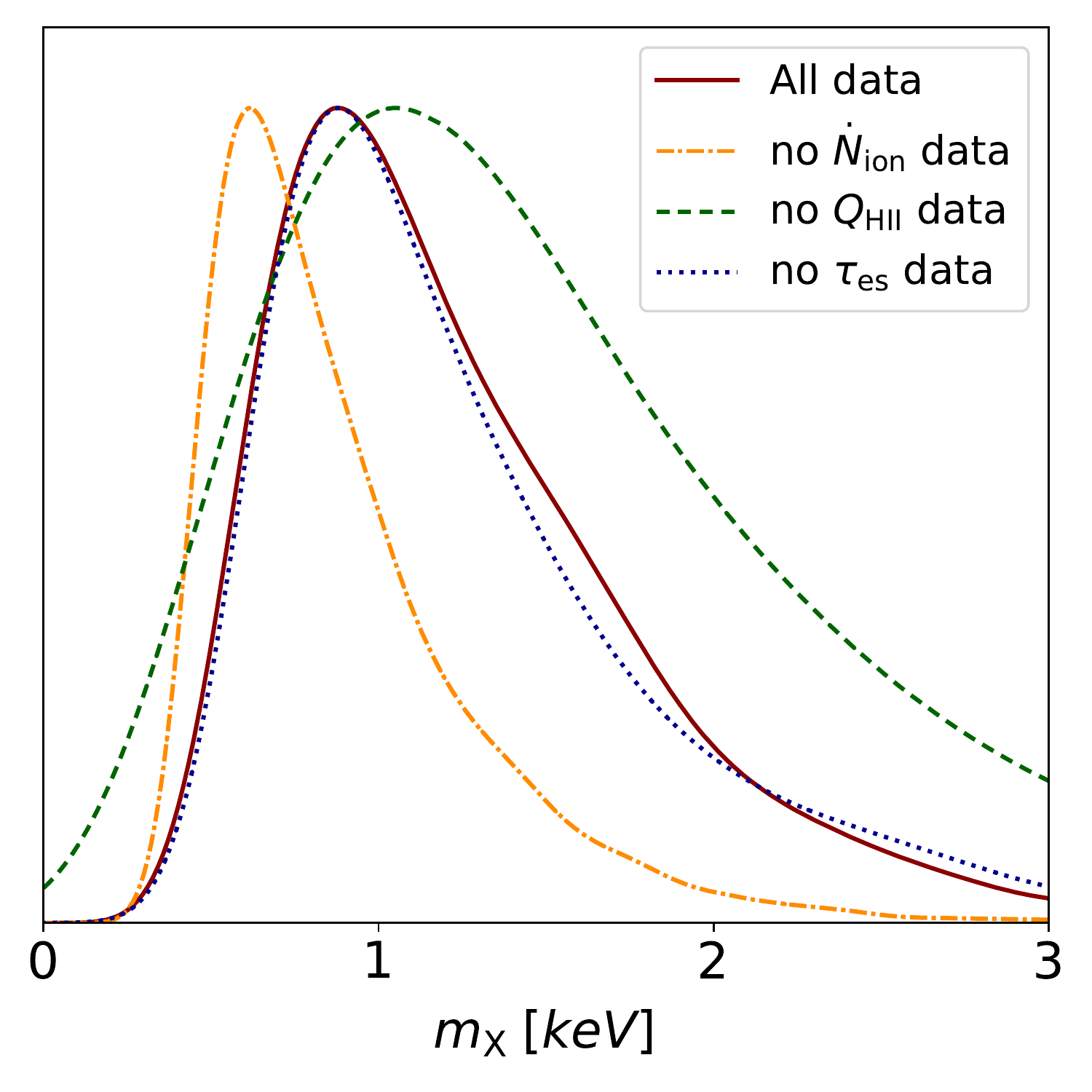}\\
\includegraphics[width=9cm,height=7cm]{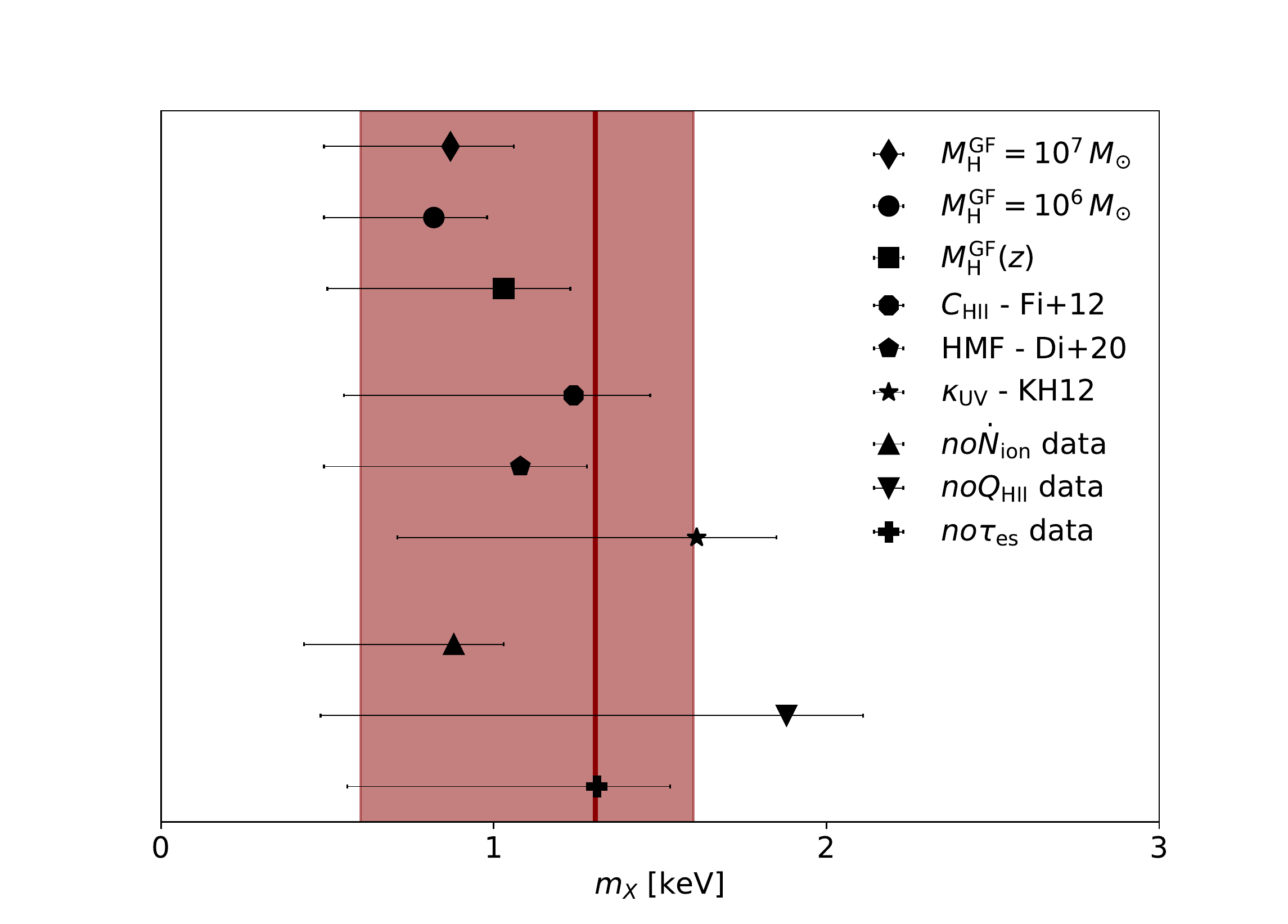}
\end{tabular}
\caption{Top panels: dependence of the posterior distributions (normalized to $1$ at their maximum value) for the DM particle's  mass $m_X$ in the WDM scenario on a few assumptions adopted in this work; in all top panels the fiducial case is illustrated as a red solid line. Top left panel: effects of changing the threshold halo mass $M_{\rm H}^{\rm GF}$ of galaxy formation from our fiducial value $10^8$ M$_\odot$ to $10^7$ M$_\odot$ (dot--dashed yellow), to $10^6$ M$_\odot$ (dashed green), and to the redshift-dependent atomic cooling limit $M_{\rm H}^{\rm GF}(z)$ (dotted blue).
Top middle panel: effects of changing the clumping factor $C_{\rm HII}$ from our fiducial expression by \cite{Pawlik09} to that by \cite{Finlator12} (dashed yellow), the CDM halo mass function from our fiducial determination by \cite{Tinker08} to that by \cite{Diemer20} (dashed green), and the UV luminosity to SFR conversion factor $\kappa_{\rm UV}$ from our fiducial value by \cite{Cai14} to the value by \cite{Kennicutt12} (dotted blue). Top right panel: effects of removing from the likelihood the dataset relative to the cosmic ionization rate $\dot N_{\rm ion}$ (dot--dashed yellow), to the evolution of the ionized fraction $Q_{\rm HII}$ (dashed green) and the optical depth $\tau_{\rm es}$ (dotted blue). Bottom panel: summary plot showing the mean and the $1\sigma$ dispersion of the posterior distributions presented in the top panels, as indicated in the legend; for reference, the outcome in our fiducial setup is illustrated by the red vertical line and shaded area.}
\label{fig|WDM_comp}
\end{figure}

Finally, to test the robustness of our astroparticle posterior estimates, we vary some of the assumptions made in our fiducial setup above, focusing on the WDM scenario for definiteness; the impact on the marginalized distribution of the DM particles' mass $m_X$ is shown in Figure~\ref{fig|WDM_comp}. In the top left panel, we change the threshold halo mass $M_{\rm H}^{\rm GF}$ of galaxy formation; instead of our fiducial value $10^8$ M$_\odot$ (red solid), we try with  $10^7$ M$_\odot$ (dot--dashed yellow) , $10^6$ M$_\odot$ (dashed green), and with the redshift-dependent atomic cooling limit $\log M_{\rm H}^{\rm GF}(z)$ [M$_\odot]\approx 8.41-0.092\times (z-4)+0.0023\times (z-4)^2$ (dotted blue; see \cite{BL01,Finkelstein19}). Such smaller values of $M_{\rm H}^{\rm GF}$ could be possibly associated with star formation in mini halos, although typically this occurs at redshifts $z\gtrsim 15-20$ higher than those considered here when setting the galaxy formation constraint. The net effect of a smaller $M_{\rm H}^{\rm GF}$ is to narrow somewhat the high-mass tail of the marginalized distribution and to shift its maximum toward slightly smaller values; the overall constraints on $m_X$ are not appreciably altered, since it can be seen from Figure~\ref{fig|abma} that for $m_X\sim$ keV the magnitude values corresponding to a given halo mass are very similar in the range $M_{\rm H}\sim 10^{6-8}$ M$_\odot$.

In the top middle panel of Figure~\ref{fig|WDM_comp}, we explore the variations of other auxiliary quantities: clumping factor $C_{\rm HII}$ entering the recombination timescale in Equation (\ref{eq|QHII}); CDM halo mass function entering in Equation (\ref{eq|HMF}); parameter $\kappa_{\rm UV}$ determining the relation between UV luminosity and SFR (see Section \ref{sec|reion}). It is beyond the scope of the paper to investigate all of the possible variations of these quantities in a systematic way, so we just focus on other choices often exploited in the literature. Specifically, with respect to our fiducial cases (red solid) we try the following: instead of the clumping factor\endnote{note that the clumping factor may actually depend on cosmology itself; in particular, as shown \mbox{by \cite{Romanello21}}, based on the model \mbox{by \cite{Trombetti14}}, for a WDM scenario in the relevant redshift range $z\gtrsim 6$, the clumping factor tends to be slightly lower than our fiducial case by \cite{Pawlik09}. We checked that adopting such a cosmology-dependent clumping factor has a minor impact on the $m_X$ posterior.} by \cite{Pawlik09,Haardt12}, we used that by \cite{Finlator12} (dot--dashed yellow); instead of the halo mass function from \cite{Tinker08}, we used that by \cite{Diemer20} (dashed green); instead of our fiducial value, we used that by \cite{Kennicutt12} for solar metallicity (dotted blue). Overall, the constraints on $m_X$ are not substantially affected by these variations.

In the top right panel of Figure~\ref{fig|WDM_comp}, we check the dependence of our results on the set of data used to construct the likelihood in Section \ref{sec|Bayes} and Table \ref{tab|data}; in particular, with respect to our fiducial case (solid red) we remove one-by-one the constraints from the cosmic ionization rate $\dot N_{\rm ion}$ (dot--dashed yellow), from the evolution of the ionized fraction $Q_{\rm HII}$ (dashed green) and from the optical depth $\tau_{\rm es}$ (dotted blue). Not surprisingly, the most stringent constraints on $m_X$ come from the data on the redshift evolution of $Q_{\rm HII}$, in absence of which the marginalized distribution would considerably widen. The data on $Q_{\rm HII}$ tend to prefer slightly lower values of $m_X$, while those on $\dot N_{\rm ion}$ tend to prefer slightly higher values; interestingly, the combinations of these datasets produce a marginalized distribution consistent with the constraints from the latest measurements of the optical depth $\tau_{\rm es}$.

In the bottom panel of Figure~\ref{fig|WDM_comp}, we present a summary plot showing the mean and the $1\sigma$ dispersion of the posterior distributions from the top panels  (as labeled in the legend).~It is seen that the outcome of our fiducial setup (illustrated by the red vertical line and shaded area) is quite robust against all the different variations in the parameters and assumptions considered above.

\section{Summary and Outlook}\label{sec|summary}

In this work, we derived astroparticle constraints for different dark matter scenarios alternative to standard cold dark matter (CDM): thermal relic warm dark matter, WDM; fuzzy dark matter, $\psi$DM; self-interacting dark matter, SIDM; sterile neutrino dark matter, $\nu$DM. For this purpose, we relied on three main ingredients: updated determinations of the high-redshift UV luminosity functions for primordial galaxies out to redshift $z\sim 10$; redshift-dependent halo mass functions in the above DM scenarios, as provided by state-of-the art numerical simulations; robust constraints on the reionization history of the Universe from recent astrophysical and cosmological datasets.

We built up an empirical model of cosmic reionization (see Section \ref{sec|reion}) characterized by two basic parameters: the escape fraction $f_{\rm esc}$ of ionizing photons from primordial galaxies, and the limiting UV magnitude $M_{\rm UV}^{\rm lim}$ down to which the extrapolated UV luminosity functions were steeply increasing.
We performed standard abundance matching of the UV luminosity function and the halo mass function (see Section \ref{sec|abma}), obtaining a relationship between UV luminosity and halo mass whose shape depended on an astroparticle quantity $X$ specific to each DM scenario. We exploited such a relation to introduce in the analysis a constraint from primordial galaxy formation, in terms of the threshold halo mass $M_{\rm H}^{\rm GF}$ above which primordial galaxies could efficiently form stars. We performed Bayesian inference on the three parameters $f_{\rm esc}$, $M_{\rm UV}^{\rm lim}$, and $X$ via an MCMC technique (see \mbox{Section \ref{sec|Bayes}} and Figures \ref{fig|CDM}--\ref{fig|nuDM}).

The marginalized posterior estimates are discussed in Section \ref{sec|results}, and summarized in Table \ref{tab|results}. As for the astroparticle property $X$, we found: WDM particle mass $m_X\approx 1.3^{+0.3}_{-0.7}$ keV, $\psi$DM particle mass $m_X\approx 2.09^{+0.09}_{-0.05}\times 10^{-22}$ eV, SIDM temperature at kinetic decoupling $T_X\approx 0.24^{+0.04}_{-0.13}$ keV, and lepton asymmetry $L_X\approx 10.7^{+1.4}_{-1.4}$ for a sterile neutrino of mass $m_X\sim 7$ keV. A comparison with literature constraints from independent observations (see Figure~\ref{fig|APP}) seems to challenge thermal WDM and $\psi$DM as viable alternatives to CDM, while there is more room for SIDM and $\nu$DM. As for the astrophysical parameters, the values of the escape fraction $f_{\rm esc}$ were found to vary from $0.05$ to $0.15$, and those of the UV limiting magnitude ranged from $-12$ to $-16$, depending on the DM scenario (see \mbox{Table \ref{tab|results}}). We performed a model comparison among the different DM scenarios, both in terms of projection of our best-fit models on the reionization observables (see Figures \ref{fig|SFR}--\ref{fig|tau_es}), and in terms of the Bayesian inference criterion; the latter indicates evidence in favor of non-CDM scenarios, though it is risky to identify a clear preference among them.

Finally, we investigated the robustness of the estimates on the astroparticle property $X$ against educated variations of uncertain astrophysical quantities (e.g., clumping factor, halo mass function, UV luminosity to SFR conversion factor), of the galaxy formation threshold $M_{\rm H}^{\rm GF}$, and of the datasets exploited to construct the likelihood in our Bayesian analysis (see Figure~\ref{fig|WDM_comp}).

\begin{figure}[H]

\includegraphics[width=11cm]{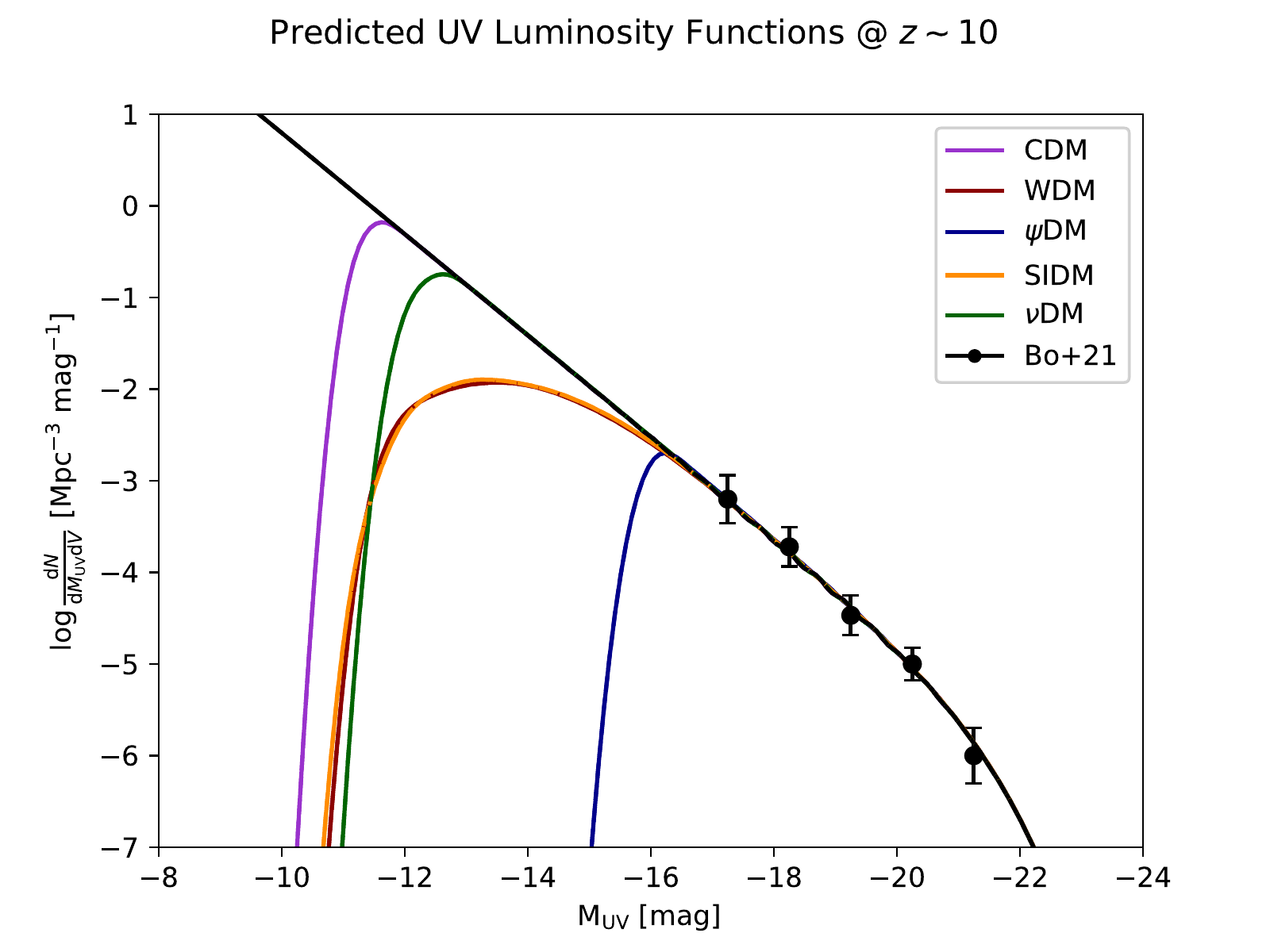}
\caption{Predicted ultra-faint end of the UV luminosity function at $z\sim 10$ in different DM scenarios: CDM (magenta), WDM (red), $\psi$DM (green), SIDM (orange), and $\nu$DM (green). Data for $M_{\rm UV}\lesssim -17$ {by} \cite{Bouwens21,Oesch18} are also illustrated for reference (black circles).}\label{fig|UVLF_pred}
\end{figure}

From a future perspective, it is worth highlighting the impacts of the different DM scenarios on the ultra-faint end of the UV luminosity function at high redshift (see \mbox{also \cite{Lapi15,Alavi16}}). We can make specific predictions by reconstructing the luminosity function from the halo mass function via
\begin{equation}\label{eq|UVLF_pred}
\frac{{\rm d}N}{{\rm d}M_{\rm UV}\,{\rm d}V} = \int{\rm d}X\, \mathcal{P}(X)\,\int_{M_{\rm H}^{\rm GF}}^{\infty}{\rm d}M_{\rm H}\, \frac{{\rm d}N_X}{{\rm d}M_{\rm H}\,{\rm d}V}\, \delta_D\left[M_{\rm UV}-M_{\rm UV}(M_{\rm H},z|X)\right]\;;
\end{equation}
where $\delta_D[\cdot]$ is a Dirac delta function centered on the inverse abundance matching relationship $M_{\rm UV}(M_{\rm H},z|X)$, $M_{\rm H}^{\rm GF}$ is the halo mass above which galaxy formation can take place, and $\mathcal{P}(X)$ is the marginalized posterior distribution of the astroparticle property $X$ specific to each DM scenario. The outcome of this computation at $z\sim 10$ is illustrated in Figure~\ref{fig|UVLF_pred}. We expect the luminosity function at the ultra-faint end to deviate from the steep behavior extrapolated from the currently observed magnitude range $M_{\rm UV}\lesssim -17$. The limiting magnitude at which the deviation occurs could be seen, and the shape of the luminosity function around that value crucially depends on the adopted DM scenario. Future observations conducted by the \textit{James Webb Space Telescope} \cite{Gardner06,Pacucci13,Atek18,Park20,Labbe21,Robertson22}, possibly eased by gravitational lensing effects, could extend the observable magnitude range down to $M_{\rm UV}\sim -13$ or fainter, thus providing valuable information  on astroparticle physics and the astrophysics of primordial galaxy formation.

\vspace{-3pt}

\funding{A.L. acknowledges funding from the EU H2020-MSCA-ITN-2019 project 860744 \textit{BiD4BESt: Big Data applications for black hole Evolution STudies} and from the PRIN MIUR 2017 prot. 20173ML3WW, \textit{Opening the ALMA window on the cosmic evolution of gas, stars, and supermassive black holes}.}

\acknowledgments{We thank the two anonymous referees for the constructive comments. We acknowledge  G. Gandolfi and P. Salucci for the stimulating discussions.}

\conflictsofinterest{The authors declare no conflict of interest.}

\begin{adjustwidth}{-\extralength}{0cm}
\reftitle{References}
\printendnotes[custom]

\end{adjustwidth}

\end{document}